\documentclass[twocolumn,prc,reprint,nofootinbib,superscriptaddress,showpacs,a4]{revtex4-1}
\usepackage{epsfig,graphicx,float,pst-all}
\usepackage{mathrsfs}
\usepackage[utf8]{inputenc}
\usepackage{rotating}
\usepackage{url}
\usepackage{esvect}
\usepackage{subfigure}
\usepackage{multirow}
\usepackage{amsmath}
\usepackage{amssymb}
\usepackage{color}

\usepackage{natbib}
\usepackage{hyperref}
\usepackage[warn]{textcomp}
\usepackage{gensymb}
\usepackage{footnote}
\usepackage{siunitx}
\usepackage{comment}
\usepackage[normalem]{ulem}
\begin{document}
\title{Exploring two-neutron halo formation in the ground-state of $\boldsymbol{^{29}}$F within a three-body model} 
\author{Jagjit Singh}
\email{jsingh@rcnp.osaka-u.ac.jp}
\affiliation{Research Center for Nuclear Physics (RCNP), Osaka University, Ibaraki 567-0047, Japan}
\author{J. Casal}
\email{casal@pd.infn.it}
\affiliation{Dipartimento di Fisica e Astronomia ``G.Galilei'', Università degli Studi di Padova, via Marzolo 8, Padova, I-35131, Italy}
\affiliation{INFN-Sezione di Padova, via Marzolo 8, Padova, I-35131, Italy}
\author{W. Horiuchi}
\affiliation{Department of Physics, Hokkaido University, Sapporo 060-0810, Japan}
\author{L. Fortunato}
\author{A. Vitturi}
\affiliation{Dipartimento di Fisica e Astronomia ``G.Galilei'', Università degli Studi di Padova, via Marzolo 8, Padova, I-35131, Italy}
\affiliation{INFN-Sezione di Padova, via Marzolo 8, Padova, I-35131, Italy}

\date{\today}

\begin{abstract}
\begin{description}
\item[Background]
The $^{29}$F system is located at the lower-$N$ boundary of the ``island of inversion'' and is an exotic, weakly bound system. Little is known about this system beyond its two-neutron separation energy ($S_{2n}$) with large uncertainties. A similar situation is found for the low-lying spectrum of its unbound binary subsystem $^{28}$F.
\item[Purpose] 
To investigate the configuration mixing, matter radius and neutron-neutron correlations in the ground-state of $^{29}$F within a three-body model, exploring the possibility of $^{29}$F to be a two-neutron halo nucleus.
\item[Method]
The $^{29}$F ground-state wave function is built within the hyperspherical formalism by using an analytical transformed harmonic oscillator basis. The Gogny-Pires-Tourreil (GPT) $nn$ interaction with central, spin-orbit and tensor terms is employed in the present calculations, together with different $\text{core}+n$ potentials constrained by the available experimental information on $^{28}$F.
\item[Results] The $^{29}$F ground-state configuration mixing and its matter radius are computed for different choices of the $^{28}$F structure and $S_{2n}$ value. The admixture of $d$-waves with $pf$ components are found to play an important role, favouring the dominance of dineutron configurations in the wave function. Our computed radii show a mild sensitivity to the $^{27}\text{F}+n$ potential and $S_{2n}$ values. The relative increase of the matter radius with respect to the $^{27}$F core lies in the range $0.1$\,-\,$0.4$\,fm depending upon these choices.
\item[Conclusions] Our three-body results for $^{29}$F indicate the presence of a moderate halo structure in its ground state, which is enhanced by larger intruder components. This finding calls for an experimental confirmation.
\end{description}
\end{abstract}
\maketitle
\section{Introduction}
The astonishing developments in the new generation of radioactive ion beam facilities have triggered many investigations to understand the shell evolution when moving away from the stability valley towards the far Eastern-region of the nuclear chart in the neutron-rich sea. The disappearance of the large shell gap at neutron number $N=20$ was reported almost four decades ago \cite{THI75,HUB78,DET79}. The small region around $N\sim20$ gained extensive attention of the nuclear physics community and is popularly known as ``island of inversion'' \cite{WAR91}. The natives of this island display exotic structural features such as dampening of shell gaps \cite{SOR08}, formation of halo and deformed structures \cite{NAKA09,MOTO95}. 

A large portion of this island is covered by the neutron-rich medium-mass isotopes of Ne, Na and Mg \cite{WAR91}. The reduced shell gap accommodates the indicative mixing of intruder $pf$-shell with the conventional $sd$-shell neutron configurations \cite{Utsuno04,Poves87,Poves02}, leading in some cases to the dominance of intruder configurations in their ground-state. Substantial efforts have been dedicated to track down the boundaries of the island of inversion and the propagation of intruder configurations along isotopic chains. For the high-$Z$ side of the island ($Z \geq 13$), the weakening of intruder configurations for $N \leq 18$ have been confirmed by $sd$ shell-model calculations \cite{Brown06}. 

The lower-$Z$ side of the island ($Z \leq 9$), however, has been relatively less explored. 
The fluorine isotopic chain provides interesting candidates to explore the extent of these intruder components in the ground state with $N \geq 19$. Little information is available on the properties of F isotopes with $N \geq 18$. Very recently, the fluorine dripline has been experimentally confirmed in $^{31}$F~\cite{Ahn19}. The $^{27}$F nucleus, sitting at the lower-$N$ border of the island, shows signatures of intruder $pf$-shell component in its excited state \cite{Elekes04}, contrary to its ground state, which is confirmed to be $sd$-shell by mass measurements \cite{Jurado07}. Also, possible indications of $pf$-intruder components in the ground state of the unbound $^{28}$F system have been reported in proton-knockout measurements \cite{CHRIS1,CHRIS2}.

For the present study, we focus on the medium-mass open shell nucleus $^{29}$F ($Z=9$ and $N=20$), whose structure is crucial for understanding the extent of the island of inversion and the shell-evolution across the F isotopic chain. The experimental and evaluated two-neutron separation energy of $^{29}$F are $S_{2n} = 1.443(436)$ MeV~\cite{GAUDE} and $1.440(650)$ MeV~ \cite{WANG}, respectively. On the theoretical side, shell-model calculations~\cite{DOOR} reported much smaller $S_{2n}$ values with large uncertainties upon the different choices of the interactions \cite{Utsuno04,Brown06,Utsuno1999}. A recent $^{29}$F measurement obtained with in-beam gamma ray spectroscopy, along with the observation of only one bound excited state, signals the necessity of neutron orbits beyond $N=20$ for a correct description of this nucleus~\cite{DOOR}.

Since $^{28}$F is neutron unbound, it is reasonable to assume that the correlation between the two valence neutrons in $^{29}$F plays an important role in binding the system. In this sense, the $^{29}$F nucleus, understood as a $^{27}$F core plus two valence neutrons, provides an example of Borromean system, such as the well studied two-neutron halo nuclei $^{6}$He or $^{11}$Li~\cite{Tanihata13}. In view of three-body models giving a good description of the structure properties of Borromean nuclei, we aim to report the first three-body calculations for the configuration mixing and matter radius of the ground state of $^{29}$F, exploring the possibility of halo formation in this region of the nuclear chart, well beyond the heaviest known two-neutron halo $^{22}$C~\cite{Horiuchi06,Togano16}. For this purpose, we solve the three-body problem in hyperspherical coordinates~\cite{Zhukov93,Nielsen01}. Here, continuum states of the two-body $^{28}$F subsystem play a crucial role, as they provide the relevant information to fix the $^{27}\text{F}+n$ potential to be used within any three-body calculations. Experimentally, the spectrum of $^{28}$F provides moderate support for a two-resonance structure with the low-lying state at $220\pm50$ keV and a high-lying resonance at $810$ keV \cite{CHRIS1,CHRIS2}, their spin-parity assignment being unclear.
This limited information will be used in the following to constrain different potential parameter sets and to investigate the sensitivity of the $^{29}$F properties to the low-lying spectrum of $^{28}$F. Given the uncertainties in the two-neutron separation energy, we will explore also its effect on the configuration mixing and matter radius, discussing different scenarios.

The paper is organized as follows. Section~\ref{MF} briefly describes the formulation of our  three-body structure model. In Sec.~\ref{28F} we analyze the subsystem $^{28}$F and fix the four different models for ${\rm core}+n$ potential, consistent with the available experimental data and theoretical predictions. Section~\ref{29F} presents our results for the three-body system $^{27}{\rm F}+n+n$, focusing on the configuration mixing in the ground state of $^{29}$F with different $S_{2n}$ values along with matter radii under different assumptions for the $^{28}$F subsystem. Finally, the main conclusions are summarized in Sec.~\ref{Sum}. 

\section{Model Formulation}
\label{MF}
Within the hyperspherical formalism~\cite{Zhukov93,Nielsen01}, the Hamiltonian eigenstates of a three-body system can be written as
\begin{equation}
  \Psi(\rho,\Omega) = \rho^{-5/2}\sum_{\beta}R_\beta(\rho)\mathcal{Y}_\beta(\Omega),
  \label{eq:3bwf}
\end{equation}
where $\rho^2=x^2+y^2$ is the hyperradius defined from the usual Jacobi coordinates $\left\{\boldsymbol{x},\boldsymbol{y}\right\}$, and $\Omega=\left\{\alpha,\widehat{x},\widehat{y}\right\}$ represents all the angular 
dependence, comprising the hyperangle $\alpha=\arctan{(x/y)}$. Note that there are three possible choices of Jacobi coordinates, although a fixed set 
will be assumed here for simplicity. 
In Eq.~(\ref{eq:3bwf}), label $\beta\equiv\{K,l_x,l_y,l,S_x,j_{ab}\}j$ is typically referred to as channel, so that $R_\beta(\rho)$ is the radial wave function for each one, and functions $\mathcal{Y}_{\beta}(\Omega)$ are states of good total angular momentum $j$ following the coupling order 
\begin{equation}
\mathcal{Y}_{\beta}(\Omega)=\left\{\left[\Upsilon_{Kl}^{l_xl_y}(\Omega)\otimes\kappa_{S_x}\right]_{j_{ab}}\otimes\phi_I\right\}_{j\mu}.
\label{eq:Upsilon}
\end{equation}
Here, $\Upsilon_{Klm_l}^{l_xl_y}(\Omega)$ are the hyperspherical harmonics~\cite{Zhukov93}, eigenstates of the hypermomentum operator $\widehat{K}$. It is then clear that $\boldsymbol{l}=\boldsymbol{l}_x+\boldsymbol{l}_y$, $S_x$ is the total spin of the two particles related by the $x$ coordinate, $\boldsymbol{j}_{ab}=\boldsymbol{l}+\boldsymbol{S}_x$, and $I$ stands for the spin of the core nucleus, which is assumed to be fixed. More details can be found, for instance, in Ref.~\cite{Casal18}.

In order to determine the radial functions for a given system, we use the pseudostate method~\cite{Tolstikhin97}, which consists in diagonalizing the Hamiltonian in a suitable basis. In this approach, in addition to negative-energy eigenstates associated to bound states, the resulting positive-energy eigenstates (or pseudostates) provide a discrete representation of the continuum. In the present work, however, we focus only on bound states. Then, radial functions will be written as
\begin{equation}
    R_\beta(\rho)=\sum_{i} C_{i\beta} U_{i\beta}(\rho),
    \label{eq:PS}
\end{equation}
where the index $i$ counts the number of basis functions, or hyperradial excitations, and $C_{i\beta}$ are just diagonalization coefficients. For this purpose, different bases can be used~\cite{Desc03,Matsumoto04,MRoGa05}, but in the present work we employ the analytical transformed harmonic oscillator (THO) basis~\cite{Casal13}. 

The diagonalization of the three-body Hamiltonian requires the computation of the corresponding kinetic energy and potential matrix elements. With the above definition in hyperspherical coordinates, we can write~\cite{CasalTh,IJThompson04}
\begin{equation}
    T_\beta(\rho)=-\frac{\hbar^2}{2m} \left(\frac{d^2}{d\rho^2}-\frac{15/4+K(K+4)}{\rho^2} \right),
    \label{eq:top}
\end{equation}
for the kinetic energy operator, where $m$ is a normalization mass, typically the mass of the nucleon, while the coupling potentials are given by
\begin{equation}
V_{\beta\beta'}(\rho)=\left\langle \mathcal{Y}_{\beta }(\Omega)\Big|V_{12}+V_{13}+V_{23} \Big|\mathcal{Y}_{\beta'}(\Omega) \right\rangle + \delta_{\beta\beta'}V_{\rm 3b}(\rho).
\label{eq:vcoupl}
\end{equation}
In this expression, $V_{ij}$ are the corresponding two-body interactions within the three-body composite, which are fixed by the known experimental information on the binary subsystems. Then, a phenomenological three-body force $V_{\rm 3b}(\rho)$ is customarily introduced to account for effects not explicitly included in a three-body picture with two-body interactions alone~\cite{IJThompson04,MRoGa05,RdDiego10,Casal13}. This term can be used as the only free parameter in this model, to adjust the energy of the states to their known experimental value, if available. Note that some authors use instead a scaling parameter or renormalization factor in the binary potentials to fix the three-body energies~\cite{Desc03,HAG,JS16}.

\section{Unbound two-body system ($\boldsymbol{^{27}{\rm F}+n}$)}
\label{28F}
The spectral properties of $\text{core}+n$ subsystems play a fundamental role in the structure of Borromean three-body nuclei, as the corresponding potential enters explicitly in the Hamiltonian through Eq.~(\ref{eq:vcoupl}). In the present case, this amounts to fixing a $^{27}\text{F}+n$ potential to describe the low-lying continuum spectrum of $^{28}$F. Though $N=18$ is open in the $1d_{3/2}$ subshell, we still model $^{28}$F as a $^{27}$F core surrounded by an unbound neutron moving in $d_{3/2}$, $f_{7/2}$ and $p_{3/2}$ orbitals in a simple independent-particle shell-model picture. While the separation between core and valence neutron is not so clear in this system, for simplicity we consider an inert-core approximation, in such a way that any possible effects coming from internal rearrangements or core-valence exchange will be somehow contained in $l$-dependent potential parameters. Note that a similar approach has been followed in other three-body calculations, for instance, the $^{14}\text{Be}+n$ subsystem for the description of $^{16}$Be~\cite{Lovell17}. Moreover, we disregard the spin of the unpaired proton in $^{27}$F, since we consider only neutron degrees of freedom. This will simplify the construction of the $^{29}$F three-body ground-state wave function as a $0^+$ state.

The only experimental study on the spectrum of the neutron unbound $^{28}$F reports the indication of two resonances at low energies ($<1$\,MeV) by making use of the invariant mass spectroscopy technique \cite{CHRIS1,CHRIS2}. However, due to experimental limitations, this does not rule out the existence of other possibilities such as a single resonant structure or more than two states, and in any case the spin-parity assignment is not clear at all.
Under this uncertain experimental situation, we construct the $^{27}\text{F}+n$ potential including central and spin-orbit terms with Woods-Saxon geometry,
\begin{equation}
V_{{\rm core}+n} = \left(-V_0+V_{ls}\lambda_\pi^2 \vec{l}\cdot\vec{s}\frac{1}{r}\frac{d}{dr}\right)f(r) ,
\label{vls}
\end{equation}
where $V_0$ can be in general $l$-dependent, and $f(r)=(1+{\rm\,exp}[(r-R_c)/a])^{-1}$, with $R_c=r_0A_c^{1/3}$ ($A_c=27$ for $^{27}$F). The spin-orbit interaction in Eq.~(\ref{vls}) is written in terms of the Compton wavelength $\lambda_\pi=1.414$ fm. By following the prescription of Ref.~\cite{HOR10}, the spin-orbit strength is set to follow the systematics \cite{BOHR} and is $V_{ls}=17.333$ MeV. The values of $r_0=1.25$ fm and $a=0.75$ fm are also adopted from the Ref.~\cite{HOR10}, originally suggested for $^{31}$Ne. 

Considering the limited experimental information, we examine four different scenarios for the present study, with the corresponding sets of potential parameters listed in Table~\ref{T1}. In Set A, $V_0$ is chosen to be $l$-independent and adjusted to fix the $d_{3/2}$ ground-state resonance of $^{28}$F at $0.20$ MeV, corresponding to the lowest peak reported in Ref.~\cite{CHRIS1}. This set follows the standard shell-model scenario, with an additional $f_{7/2}$ resonance appearing at relatively higher energy, $\sim 3$\,MeV.  
\begin{table}
\caption{Parameter sets for the $^{27}$F$+n$ interaction, where $V_{0}$ is the Woods-Saxon potential depth, $E_{R}$ is the position of the resonances and $\Gamma$ is their width. Note that $r_0=1.25$ fm, $a=0.75$ fm and $V_{ls}=17.333$ MeV are fixed.}
\centering
\begin{tabular}{ccccc}
\toprule\\[-1.5ex]
SET & $lj$      & $V_0$\,(MeV) & $E_{R}$\,(MeV) & $\Gamma$\,(MeV)  \\ [1ex]
\colrule\\[-1.5ex]
             & $d_{3/2}$ & 39.67         & 0.20         & 0.007   \\
 \textbf{A}  & $p_{3/2}$ & 39.67         & ...          & ...     \\
             & $f_{7/2}$ & 39.67         & 3.36         & 0.638   \\ [1ex]
\colrule\\[-1.5ex]
             & $d_{3/2}$  & 39.67        & 0.20         & 0.007   \\
 \textbf{B}  & $p_{3/2}$  & 44.80        & 0.80         & 3.979   \\
             & $f_{7/2}$  & 44.80        & 1.67         & 0.074   \\ [1ex]
\colrule\\[-1.5ex]
             & $d_{3/2}$  & 39.67        & 0.20         & 0.007   \\
 \textbf{C}  & $p_{3/2}$  & 46.78        & 0.20         & 0.199   \\
             & $f_{7/2}$  & 39.67        & 3.36         & 0.638   \\ [1ex]
\colrule\\[-1.5ex]
             & $d_{3/2}$  & 30.00        & 4.70          & 8.818     \\
 \textbf{D}  & $p_{3/2}$  & 46.78        & 0.20          & 0.199   \\
             & $f_{7/2}$  & 30.00        & 6.96          & 5.714     \\ [1ex]
\botrule
\end{tabular}
\label{T1}
\end{table}
\begin{figure}
\centering
\includegraphics[width=\linewidth]{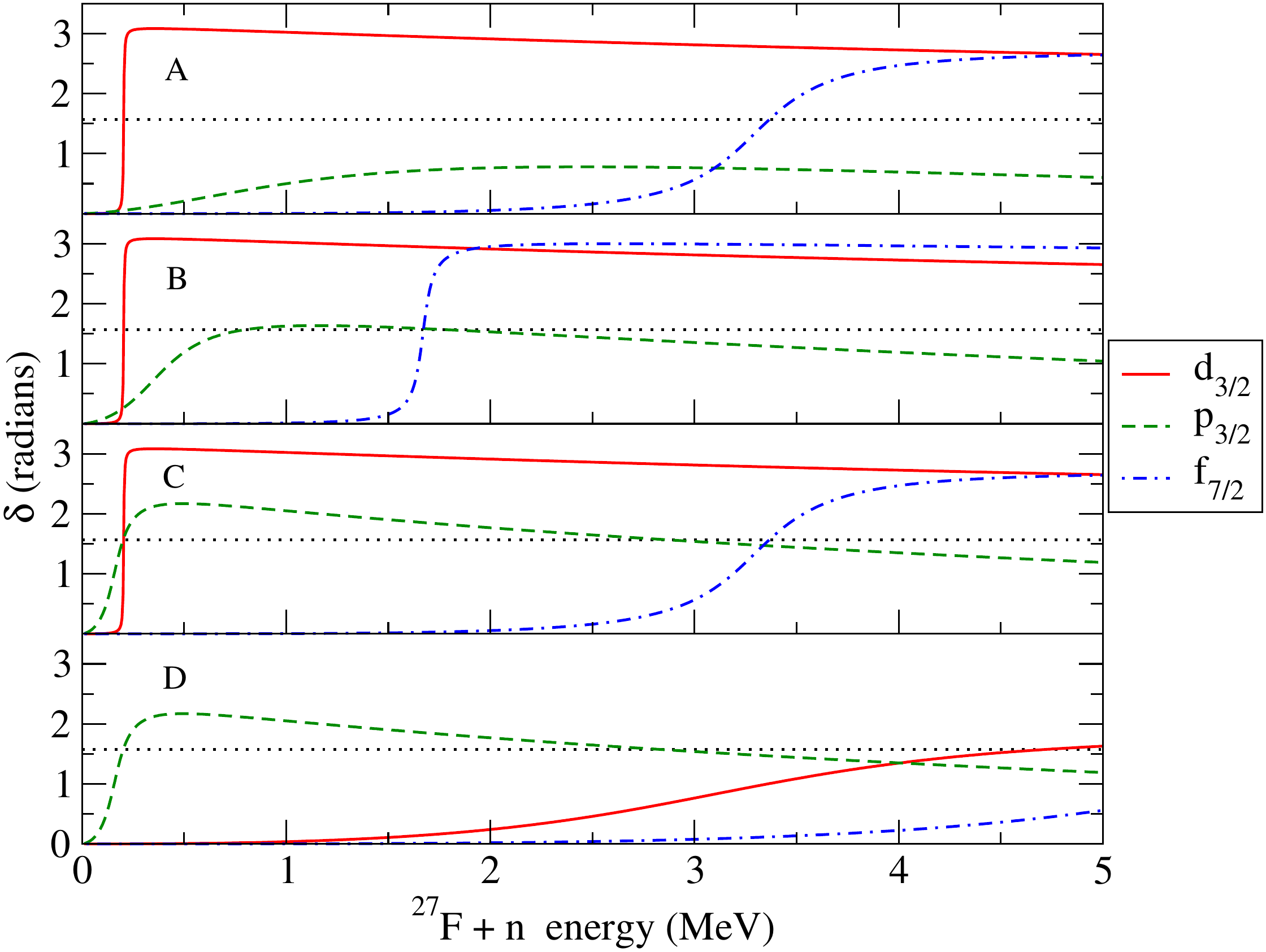}
\caption{(Color online) $^{27}{\rm F}+n$ phase shifts for $d_{3/2}$, $p_{3/2}$ and $f_{7/2}$ states, corresponding  to different Sets (A-D). The dotted black line corresponds to $\pi/2$.}
\label{FIG1}
\end{figure}

In order to explore the interference of $pf$-shell with conventional $sd$-shell, we consider an intruder scenario defined by Set B. In this case we have tuned the $p$-wave strength so that the $p_{3/2}$ resonance coincides with the position of the second peak reported in Ref.~\cite{CHRIS1}. For this set, we use the same depth ($V_0$) for the negative-parity states, which pushes the $f_{7/2}$ resonance to a lower energy with respect to Set A.

For neutron-rich Ne isotopes, Monte Carlo shell-model calculations suggest that the ${3/2}^+$ and ${3/2}^-$ states are practically degenerate, pointing towards extreme gap quenching between $d_{3/2}$ and $p_{3/2}$~\cite{Koba2016}. We consider the possibility of a degenerate scenario for the F isotopic chain by introducing Set C, in which we change only the $p$-wave strength with respect to Set A.

Finally, we propose an additional, extreme inverted scenario assuming the ground state of $^{27}$F to be the $p$-wave resonance, while the $d$- and $f$-states are pushed to higher energies. This set is included to later explore the increment of $p$-wave content in the ground state of the $^{29}$F three-body system, which may significantly affect the corresponding radius and the discussion regarding the possibility of halo formation in this nucleus.

The phase shifts corresponding to $d_{3/2}$, $p_{3/2}$ and $f_{7/2}$ states of $^{28}$F for all potential sets A-D are plotted in Fig.~\ref{FIG1}. The position and widths of these possible resonances are tabulated in Table~\ref{T1}. In addition, our $\text{core}+n$ potentials produce $1s_{1/2}$, $1p_{3/2}$, $1p_{1/2}$, $1d_{5/2}$ and $2s_{1/2}$ bound states which represent the fully occupied neutron orbitals of the core. Note that we use, for $s$ waves, the potential depth $V_0=39.67$\,MeV in all sets, while the potential for the $p_{1/2}$ state is always the same as the one employed for its $p_{3/2}$ partner in each case. These Pauli forbidden states would give rise to unphysical eigenstates of the three-body Hamiltonian, so they need to be removed for the three-body calculations discussed in the next section. In this work, this is achieved by means of a supersymmetric transformation~\cite{Baye87,Baye287}, which yields spectrally equivalent potentials without the bound states.

\section{Three-body ($\boldsymbol{^{27}{\rm F}+n+n}$) calculations}
\label{29F}
Using the $n+{^{27}}$F potentials described in the previous section, we compute the three-body ground-state of the $^{29}$F nucleus considering that the two valence neutrons couple to $0^+$.  For that purpose, we need also the $nn$ interaction. In this work, we adopt the Gogny-Pires-Tourreil (GPT) potential~\cite{Gogny70} including central, spin-orbit and tensor terms, as in Refs.~\cite{Lovell17,IJThompson04,Casal13} and several other three-body calculations in the literature. Then, as already introduced by Eq.~(\ref{eq:vcoupl}), a phenomenological three-body force is employed to fix the energy of the state. This force can be modeled as a simple Gaussian potential~\cite{CasalTh},
\begin{equation}
    V_{3b}(\rho)=v_{3b}e^{-{\left(\rho/\rho_o\right)}^2},
    \label{eq:3bforce}
\end{equation}
where $\rho_o=6$ fm and the strength $v_{3b}$ is adjusted to recover different $S_{2n}$ values. In all the cases considered, $|v_{3b}|< 6$ MeV. The three-body Hamiltonian so obtained is diagonalized in a THO basis including angular components up to a maximum hypermomentum $K_{max}$ in the wave-function expansion (Eq.~(\ref{eq:3bwf})) and $i=0,\dots,N$ excitations for the hyperradial functions (Eq.~(\ref{eq:PS})). This is done in the so-called Jacobi-T coordinate system in which the $x$ coordinate relates the two neutrons. Note that, once $K_{max}$ is fixed, the orbital angular momenta associated to each Jacobi coordinate are restricted to $l_x+l_y \le K$~\cite{Zhukov93}, and no additional truncation is needed. In this work, $K_{max}=30$ and $N=20$ are found to provide converged results.

It is worth noting that the particular choice of the $nn$ interaction is not very important for the ground-state properties of $\text{core}+n+n$ systems, as discussed in Ref.~\cite{Zhukov93}, provided the interaction describes $nn$ scattering data reasonably. We have checked this explicitly by employing a different parametrization, such as the potential given in Refs.~\cite{Garrido97,Garrido04}. This nucleon-nucleon interaction provides less binding, so that slightly deeper three-body potential strengths are needed to recover the same three-body energies. Nevertheless, the resulting wave functions are practically indistinguishable, in terms of radii and partial-wave content, from those obtained using the GPT potential.

In order to compute the $^{29}$F matter radius within the present three-body model~\cite{CasalTh}, i.e.,
\begin{equation}
 R_m = \sqrt{\frac{1}{A}\left(A_c R^2_c + \langle \rho^2\rangle\right)}. 
 \label{eq:radius}
\end{equation}
the size of the $^{27}$F core is required as input. We have used $R_c=R_m(^{27}$F$)=3.218$ fm, obtained from a $^{25}$F$+n+n$ calculation using the $^{25}$F$+n$ potential by Hagino and Sagawa \cite{HAG}, fixing $S_{2n}$($^{27}$F) to the experimental value by Gaudefroy {\it et al.}~\cite{GAUDE} and adopting the experimental value for core ($^{25}$F) matter radius from Ref.~\cite{Ozawa01}.
With this prescription, our core radius is somewhat consistent with the absorption radius extracted in Ref.~\cite{KHO}. 
However, given the uncertainties in the experimental values, our computed $^{29}$F radii must be analyzed in relative terms, i.e., by 
comparing the results for different potential models and the relative increase between $^{27}$F and $^{29}$F.

\begin{table}
\caption{The contribution of different configurations (in \%) and radial properties for the ground state of $^{29}$F, with $S_{2n}=1.440$ MeV, corresponding to each model (A-D). $R_m$ is the matter radius, while $r_{nn}$ and $r_{c-nn}$ are the root mean square distance between the valence neutrons and that of their center of mass with respect to the core, respectively (in fm).}
\centering
\begin{tabular}{ccccccc}
\toprule\\[-1.5ex]
SET    &$(d_{3/2})^2$  &$(f_{7/2})^2$ &$(p_{3/2})^2$  &$R_m$  &$r_{nn}$  &$r_{c-nn}$\\[1ex]
\colrule\\[-1.5ex]
\textbf{A}&81.3        &8.4             &6.8           &3.323       & 5.476         &3.702\\
\textbf{B}&50.7        &21.1             &21.6           &3.347       & 5.322          &4.077 \\
\textbf{C}&45.4	&7.4	&39.8	&3.380&				5.756&	4.338\\
\textbf{D}&4.2	 &2.1	&85.4	&3.459&7.210	&4.694\\[1ex]
\botrule
\end{tabular}
\label{T2}
\end{table}

With all these ingredients, our three-body results for the ground state of $^{29}$F using different choices of the $\text{core}+n$ potential (A-D) are shown in Table~\ref{T2}. In these calculations, $S_{2n}$ has been fixed to the experimental value of 1.440 MeV \cite{GAUDE,WANG}. We report the partial wave content corresponding to the two valence neutrons occupying different single-particle states, together with the radial properties. To extract the former, the wave function following the couplings given by Eqs.~(\ref{eq:3bwf}) and (\ref{eq:Upsilon}) needs to be rewritten in an appropriate way. To that aim we perform first a transformation to the Jacobi-Y set in which the $x$ coordinate connects the core with a single neutron. This representation is more similar to a typical shell-model picture, and it can be easily achieved from the Reynal-Revai coefficients~\cite{RR70,IJThompson04}. Subsequently, we change the coupling order in terms of a single-particle angular momentum $\boldsymbol{J}=\boldsymbol{l}_x+\boldsymbol{S}_x$. Note that $S_x$ is just the spin of a single neutron in the Jacobi-Y set, provided we assume the core to have no spin. 

As it can be clearly seen from Table~\ref{T2}, when moving from Set A to D the $p$-wave content in the ground state increases, which leads to larger values for the matter radius. The results for Set A shows the dominance of normal shell-model $d$-wave component in the ground state leading to the smallest radius. On the other hand, for Sets B and C our results show mixing of intruder $p$- and $f$-wave components with the normal $d$-wave configuration, as expected from the lower relative position of the negative-parity resonances when using these potentials, while Set D leads to $p$-wave intruder dominance in the ground state. We can look into the wave function in more detail by analyzing the corresponding probability densities in the Jacobi-T system, i.e., as a function of the distance between the valence neutrons ($r_{nn}$) and that between the center of mass of the neutrons and the core ($r_{c-nn}$). This is given by~\cite{Zhukov93}
\begin{equation}
    P(x,y)=x^2y^2\int d\widehat{x}d\widehat{y} \left|\Psi(\boldsymbol{x},\boldsymbol{y})\right|^2,
    \label{eq:prob}
\end{equation}
Note that the relation between scaled Jacobi coordinates and the physical distances for a $\text{core}+n+n$ system in the Jacobi-T set is~\cite{Zhukov93}
\begin{align}
    \boldsymbol{x} & =\boldsymbol{r}_{nn}\sqrt{\frac{1}{2}}, \\
    \boldsymbol{y} & =\boldsymbol{r}_{c\text{-}nn}\sqrt{\frac{2A_c}{A}}.
\end{align}
\begin{figure}[h]
    \centering
    \includegraphics[width=1.0\linewidth]{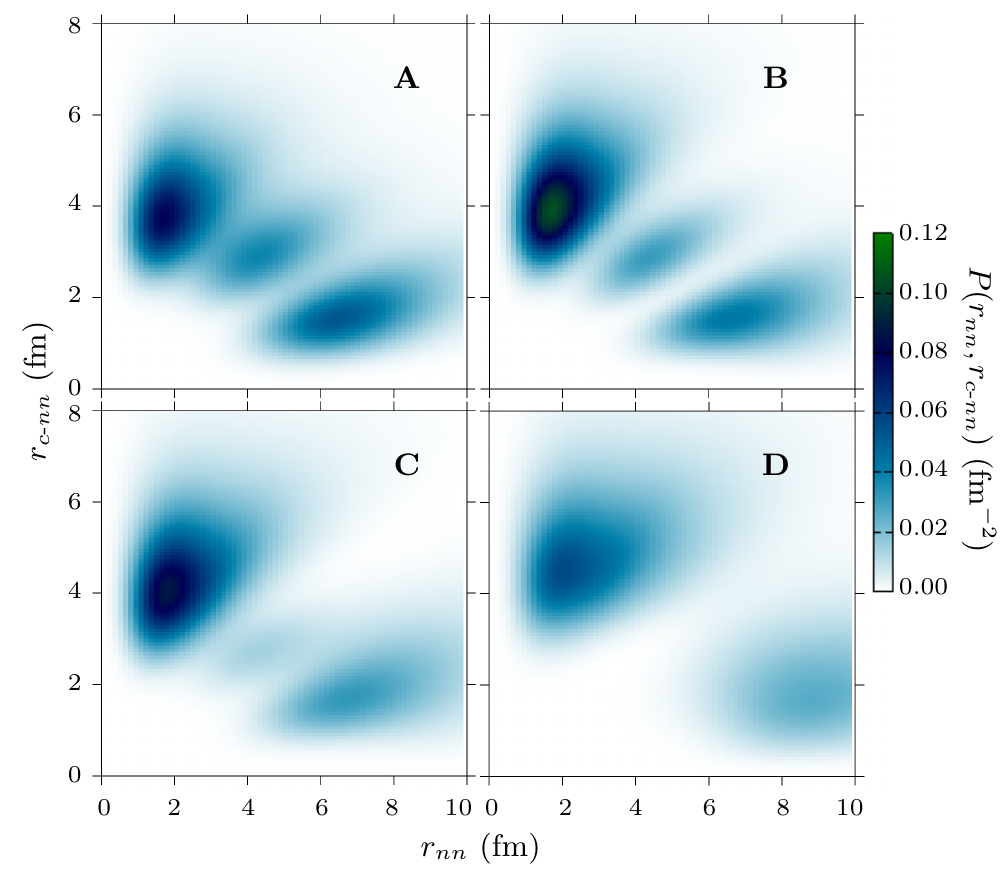}
    \caption{Ground-state probability density of $^{29}$F using the four $n$-$^{27}$F potential sets (A-D), as a function of $r_{nn}$ and $r_{c\text{-}nn}$ with the same scale (in fm$^{-2}$). In all cases, the two-neutron separation energy has been adjusted to $S_{2n}=1.440$ MeV. }
    \label{fig:prob}
\end{figure}

Our density distributions according to these definitions are shown in Fig.~\ref{fig:prob}. Note that the number of maxima in these plots is related to the dominant wave-function components, so that three peaks reflect a significant $d$-wave contribution. It is apparent that the so-called dineutron configuration, i.e., the peak corresponding to two-neutrons close to each other at some distance from the core, is dominant and more localized for sets B and C. The dineutron maxima in these models accumulate about twice as much probability as the opposite ``cigar''-like peak, in contrast to sets A and D in which this ratio is closer to unity. This is a consequence of a larger mixing between different-parity states when using sets B and C~\cite{Catara84}, and resembles the case of the two-neutron halo nuclei $^{11}$Li~\cite{sag15} or $^{14}$Be~\cite{corsi19}. With the adopted $S_{2n}$ value, however, the spatial extension of the wave function is not as large, with respect to the core, as expected for a typical halo nucleus.

\begin{table}
\caption{The contribution of different configurations (in \%) and matter radius for the ground state of $^{29}$F for each model, corresponding to four different $S_{2n}$ values (see text for details). $\Delta R$ ($=R_m-R_{\rm c}$) is the change in radius with respect to the matter radius of the core.}
\centering
\begin{tabular}{ccccccc}
\toprule\\[-1.5ex]
SET &$S_{2n}$\,(MeV)    &$(d_{3/2})^2$  &$(f_{7/2})^2$ &$(p_{3/2})^2$  &$R_m$\,(fm)&$\Delta R$\,(fm)   \\[1ex]
\colrule\\[-2.0ex]
          &0.400    &78.7	       &8.1             &9.0	       &3.363   & 0.145          \\     
          &0.790    &80.0        &8.3             &7.9           &3.343   & 0.125       \\
\textbf{A}&1.440    &81.3        &8.4             &6.8           &3.323   & 0.105     \\
	      &2.090    &82.3	       &8.5	            &6.0           &3.311   & 0.093       \\[1ex]
\colrule\\[-2.0ex]
\\ [-2.0ex]
          &0.400    &43.2	       &18.0	         &30.8	         &3.420     &   0.202   \\
          &0.790    &46.8        &19.6             &26.2	         &3.380     &   0.162    \\
\textbf{B}&1.440    &50.7        &21.1             &21.6           &3.347     &   0.129     \\
	      &2.090    &53.4	       &22.0             &18.5           &3.329     &   0.111\\[1ex]
\colrule\\[-2.0ex]
\\ [-2.0ex]
        &0.400    &30.3	&5.6	&55.7	&3.507&0.289\\
	&0.790    &37.3	&6.5	&48.2	&3.434&0.216\\
\textbf{C}&1.440    &45.4	&7.4	&39.8	&3.380&0.162\\
	&2.090    &51.4	&8.0	&33.9	&3.352&0.134\\[1ex]
\colrule\\[-2.0ex]
\\ [-2.0ex]
        &0.400    &2.8	 &1.5	&87.6	&3.598&0.380\\
	&0.790    &3.4	 &1.7	&86.7	&3.520&0.302\\
\textbf{D}&1.440    &4.2	 &2.1	&85.4	&3.459&0.241\\
	&2.090    &5.0	 &2.3	&84.2	&3.425&0.207\\[1ex]
\botrule
\end{tabular}
\label{T3}
\end{table}
\begin{figure}
\centering
\includegraphics[width=0.9\linewidth]{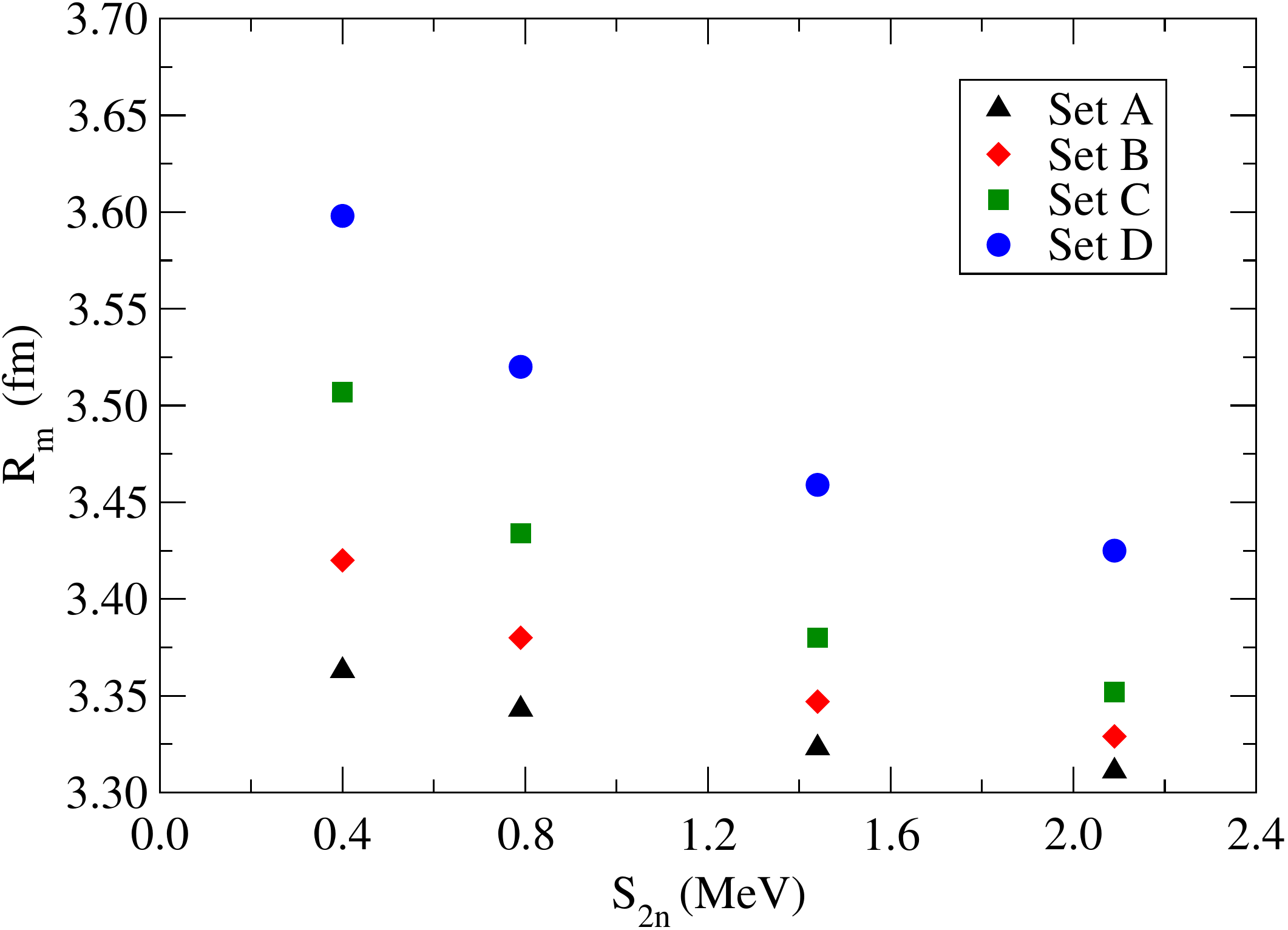}
\caption{(Color online) Matter radius ($R_m$) for the ground state of the $^{29}$F as a function of $S_{2n}$, for different sets (A-D).}
\label{FIG3}
\end{figure}

As already discussed, there are large uncertainties in the experimental and evaluated $S_{2n}$ value \cite{GAUDE,WANG}, and also shell-model calculations predict much lower values (for details see Table I of Ref.~\cite{DOOR}). 
In view of these uncertainties, we explore the sensitivity of the configuration mixing and matter radius of the ground state with $S_{2n}$ by performing additional calculations fixing the energy to the upper and lower limits of the experimental value, $2.090$ and $0.790$ MeV. Additionally, we consider also a shallower case fixing  $S_{2n}=0.400$ MeV, in accord with some of the theoretical predictions. The computed partial wave content and radii are shown in Table~\ref{T3}, together with the central values already presented in Table~\ref{T2}. As expected, a much shallower ground state yields a larger radius, as shown in Fig.~\ref{FIG3}. However, it should be noted that the calculated radii are similar among the different scenarios considered, with differences just within a $8\%$ variation, so distinguishing between them may pose an experimental challenge. New and more precise experimental data on the two-neutron separation energy of $^{29}$F and on the low-lying spectrum of $^{28}$F are needed to better constrain the theoretical models and to discriminate between the different wave functions here presented. In addition, data on knockout or transfer reactions, sensitive to the partial wave content of the $^{29}$F, are definitively desirable. 
\begin{figure}
\centering
\includegraphics[width=0.9\linewidth]{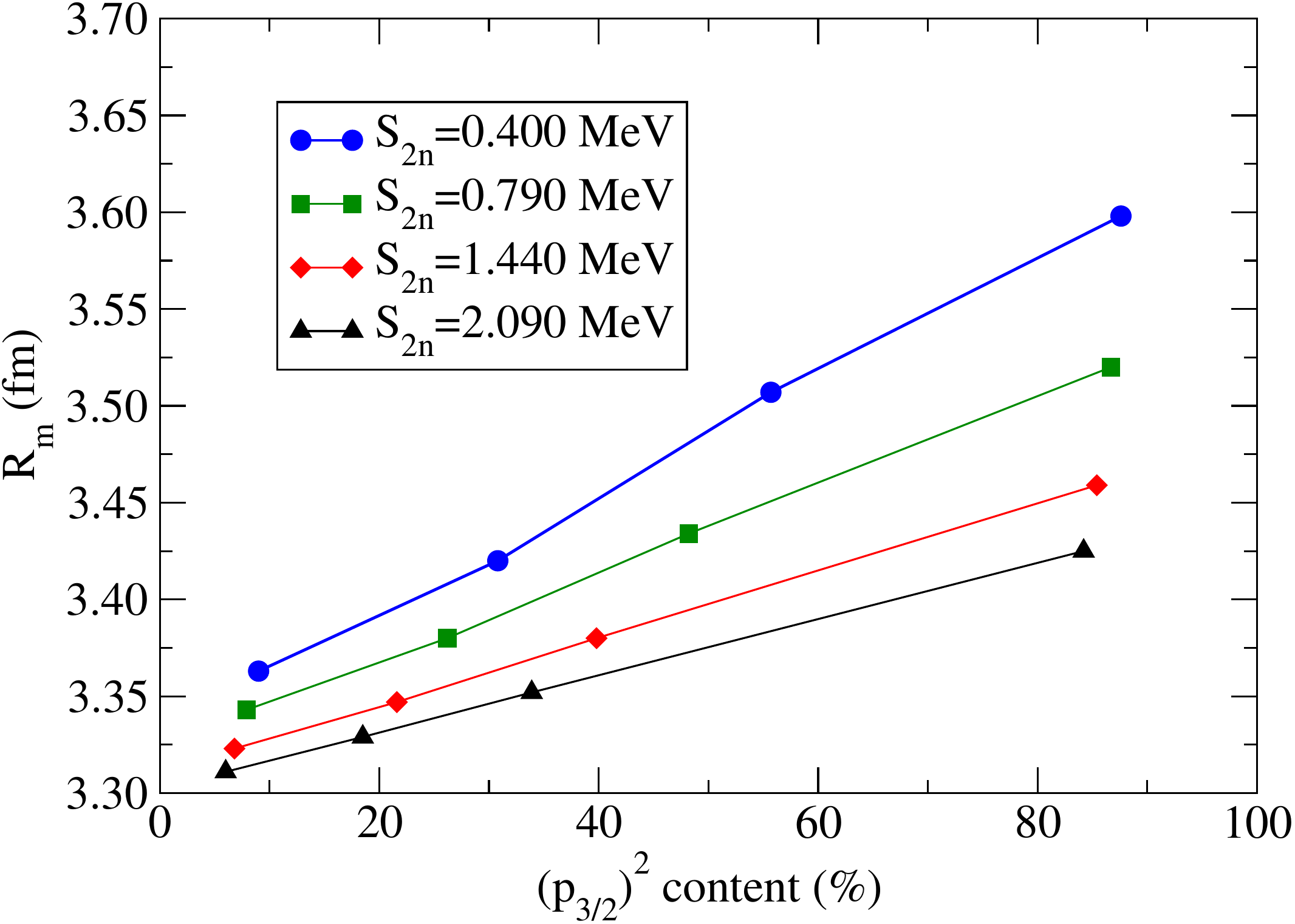}
\caption{(Color online) Matter radius ($R_m$) for the ground state of the $^{29}$F as a function of the $(p_{3/2})^2$ content for different $S_{2n}$ values.}
\label{FIG4}
\end{figure}

In Table~\ref{T3} we also give the relative change in matter radius of $^{29}$F with respect to radius of the $^{27}$F core, $\Delta R = R_m-R_c$. With the current uncertainties, the relative difference of the matter radius with respect to the $^{27}$F core ranges between $0.1$~-~$0.4$\,fm for different choices of potential sets and $S_{2n}$ values. This number is much smaller than the corresponding ones for the well-established halo nuclei $^{6}$He or $^{11}$Li. The situation for the heaviest known two-neutron halo, $^{22}$C, involves an increase of $0.45$\,fm with respect to the $^{20}$C core~\cite{Togano16}, which is likely just a consequence of the valence neutrons laying on a higher shell. In a simple estimation following the standard scaling of the radius through $A^{1/3}$, for the present $^{29}$F case we would expect $\Delta R\simeq 0.080$\,fm starting from the adopted core radius. Our three-body calculations show a mild enhancement with respect to this number, so we can tentatively conclude that the present study gives support for a moderate halo structure in the ground state of the $^{29}$F. Such a statement needs experimental confirmation in interaction cross-section measurements. The larger $\Delta R$ values correspond to the cases in which the wave function contains a significant $(p_{3/2})^2$ weight. As we show in Fig.~\ref{FIG4}, the radius scales almost linearly with the $p$-wave content, pointing toward the necessity of intruder configurations to sustain halo formation in this nucleus.

\section{Summary and Conclusions}\label{Sum}
 
We reported the first three-body (${^{27}}\text{F}+n+n$) results for the configuration mixing and matter radius of the ground state of $^{29}$F, giving special emphasis on dineutron correlations and the possibility of two-neutron halo formation. Considering the scarce available information on the unbound $^{28}$F nucleus, we conceived four different scenarios for the low-lying spectrum of this subsystem. Then, we solved the three-body problem within the hyperspherical formalism using the analytical THO basis, exploring also the uncertainties in the $S_{2n}$ value of $^{29}$F.

Our results indicate mild to strong mixing of intruder $pf$-shell configurations with normal $sd$-shell components for different choices of the $\text{core}+n$ potential. This mixing enhances the dineutron configuration in the ground-state density. The computed matter radii are within a 8\% variation depending on the different choices of the $^{27}$F$+n$ interaction and $S_{2n}$ values. This calls for new precise mass measurements and for a detailed understanding of the low-lying continuum spectrum of $^{28}$F in order to better constrain the theoretical models. Additional transfer or knockout data, sensitive to the partial wave content of $^{29}$F, could certainly help in discriminating among the different wave functions obtained in the present work.

The relative increase of matter radii with respect to $^{27}$F core lies in the range $0.1$\,-\,$0.4$\,fm in the different cases considered, which provides support for a moderate halo structure in the ground state of $^{29}$F. The radius is found to be proportional to the $(p_{3/2})^2$ content of the wave function, pointing out the relevance of intruder components in the development of the halo. Nevertheless this conclusion needs experimental confirmation via interaction cross-section measurements.

\begin{acknowledgments}
JS gratefully acknowledges the encouragement by Prof.~K.~Ogata and financial support from research budget of RCNP theory group to attend 
EFB24 2019, where this work was initiated. This work has been supported by SID funds 2019 (Investimento Strategico di Dipartimento, Università degli Studi di Padova, Italy) 
under project No.~CASA\_SID19\_1, by the Ministerio de Ciencia, Innovación y Universidades and FEDER funds (Spain) under project No.~FIS2017-88410-P, by the European Union's 
Horizon 2020 research and innovation program under grant agreement No.~654002, and by JSPS KAKENHI Grants No.~18K03635, 18H04569, and 19H05140. 
WH acknowledges the collaborative research program 2019, information initiative center, Hokkaido University.
\end{acknowledgments}

\newpage
\bibliography{ref.bib}

\begin{thebibliography}{51}%
\makeatletter
\providecommand \@ifxundefined [1]{%
 \@ifx{#1\undefined}
}%
\providecommand \@ifnum [1]{%
 \ifnum #1\expandafter \@firstoftwo
 \else \expandafter \@secondoftwo
 \fi
}%
\providecommand \@ifx [1]{%
 \ifx #1\expandafter \@firstoftwo
 \else \expandafter \@secondoftwo
 \fi
}%
\providecommand \natexlab [1]{#1}%
\providecommand \enquote  [1]{``#1''}%
\providecommand \bibnamefont  [1]{#1}%
\providecommand \bibfnamefont [1]{#1}%
\providecommand \citenamefont [1]{#1}%
\providecommand \href@noop [0]{\@secondoftwo}%
\providecommand \href [0]{\begingroup \@sanitize@url \@href}%
\providecommand \@href[1]{\@@startlink{#1}\@@href}%
\providecommand \@@href[1]{\endgroup#1\@@endlink}%
\providecommand \@sanitize@url [0]{\catcode `\\12\catcode `\$12\catcode
  `\&12\catcode `\#12\catcode `\^12\catcode `\_12\catcode `\%12\relax}%
\providecommand \@@startlink[1]{}%
\providecommand \@@endlink[0]{}%
\providecommand \url  [0]{\begingroup\@sanitize@url \@url }%
\providecommand \@url [1]{\endgroup\@href {#1}{\urlprefix }}%
\providecommand \urlprefix  [0]{URL }%
\providecommand \Eprint [0]{\href }%
\providecommand \doibase [0]{http://dx.doi.org/}%
\providecommand \selectlanguage [0]{\@gobble}%
\providecommand \bibinfo  [0]{\@secondoftwo}%
\providecommand \bibfield  [0]{\@secondoftwo}%
\providecommand \translation [1]{[#1]}%
\providecommand \BibitemOpen [0]{}%
\providecommand \bibitemStop [0]{}%
\providecommand \bibitemNoStop [0]{.\EOS\space}%
\providecommand \EOS [0]{\spacefactor3000\relax}%
\providecommand \BibitemShut  [1]{\csname bibitem#1\endcsname}%
\let\auto@bib@innerbib\@empty
\bibitem [{\citenamefont {Thibault}\ \emph {et~al.}(1975)\citenamefont
  {Thibault}, \citenamefont {Klapisch}, \citenamefont {Rigaud}, \citenamefont
  {Poskanzer}, \citenamefont {Prieels}, \citenamefont {Lessard},\ and\
  \citenamefont {Reisdorf}}]{THI75}%
  \BibitemOpen
  \bibfield  {author} {\bibinfo {author} {\bibfnamefont {C.}~\bibnamefont
  {Thibault}}, \bibinfo {author} {\bibfnamefont {R.}~\bibnamefont {Klapisch}},
  \bibinfo {author} {\bibfnamefont {C.}~\bibnamefont {Rigaud}}, \bibinfo
  {author} {\bibfnamefont {A.~M.}\ \bibnamefont {Poskanzer}}, \bibinfo {author}
  {\bibfnamefont {R.}~\bibnamefont {Prieels}}, \bibinfo {author} {\bibfnamefont
  {L.}~\bibnamefont {Lessard}}, \ and\ \bibinfo {author} {\bibfnamefont
  {W.}~\bibnamefont {Reisdorf}},\ }\href {\doibase 10.1103/PhysRevC.12.644}
  {\bibfield  {journal} {\bibinfo  {journal} {Phys. Rev. C}\ }\textbf {\bibinfo
  {volume} {12}},\ \bibinfo {pages} {644} (\bibinfo {year} {1975})}\BibitemShut
  {NoStop}%
\bibitem [{\citenamefont {Huber}\ \emph {et~al.}(1978)\citenamefont {Huber},
  \citenamefont {Touchard}, \citenamefont {B\"uttgenbach}, \citenamefont
  {Thibault}, \citenamefont {Klapisch}, \citenamefont {Duong}, \citenamefont
  {Liberman}, \citenamefont {Pinard}, \citenamefont {Vialle}, \citenamefont
  {Juncar},\ and\ \citenamefont {Jacquinot}}]{HUB78}%
  \BibitemOpen
  \bibfield  {author} {\bibinfo {author} {\bibfnamefont {G.}~\bibnamefont
  {Huber}}, \bibinfo {author} {\bibfnamefont {F.}~\bibnamefont {Touchard}},
  \bibinfo {author} {\bibfnamefont {S.}~\bibnamefont {B\"uttgenbach}}, \bibinfo
  {author} {\bibfnamefont {C.}~\bibnamefont {Thibault}}, \bibinfo {author}
  {\bibfnamefont {R.}~\bibnamefont {Klapisch}}, \bibinfo {author}
  {\bibfnamefont {H.~T.}\ \bibnamefont {Duong}}, \bibinfo {author}
  {\bibfnamefont {S.}~\bibnamefont {Liberman}}, \bibinfo {author}
  {\bibfnamefont {J.}~\bibnamefont {Pinard}}, \bibinfo {author} {\bibfnamefont
  {J.~L.}\ \bibnamefont {Vialle}}, \bibinfo {author} {\bibfnamefont
  {P.}~\bibnamefont {Juncar}}, \ and\ \bibinfo {author} {\bibfnamefont
  {P.}~\bibnamefont {Jacquinot}},\ }\href {\doibase 10.1103/PhysRevC.18.2342}
  {\bibfield  {journal} {\bibinfo  {journal} {Phys. Rev. C}\ }\textbf {\bibinfo
  {volume} {18}},\ \bibinfo {pages} {2342} (\bibinfo {year}
  {1978})}\BibitemShut {NoStop}%
\bibitem [{\citenamefont {D\'etraz}\ \emph {et~al.}(1979)\citenamefont
  {D\'etraz}, \citenamefont {Guillemaud}, \citenamefont {Huber}, \citenamefont
  {Klapisch}, \citenamefont {Langevin}, \citenamefont {Naulin}, \citenamefont
  {Thibault}, \citenamefont {Carraz},\ and\ \citenamefont {Touchard}}]{DET79}%
  \BibitemOpen
  \bibfield  {author} {\bibinfo {author} {\bibfnamefont {C.}~\bibnamefont
  {D\'etraz}}, \bibinfo {author} {\bibfnamefont {D.}~\bibnamefont
  {Guillemaud}}, \bibinfo {author} {\bibfnamefont {G.}~\bibnamefont {Huber}},
  \bibinfo {author} {\bibfnamefont {R.}~\bibnamefont {Klapisch}}, \bibinfo
  {author} {\bibfnamefont {M.}~\bibnamefont {Langevin}}, \bibinfo {author}
  {\bibfnamefont {F.}~\bibnamefont {Naulin}}, \bibinfo {author} {\bibfnamefont
  {C.}~\bibnamefont {Thibault}}, \bibinfo {author} {\bibfnamefont {L.~C.}\
  \bibnamefont {Carraz}}, \ and\ \bibinfo {author} {\bibfnamefont
  {F.}~\bibnamefont {Touchard}},\ }\href {\doibase 10.1103/PhysRevC.19.164}
  {\bibfield  {journal} {\bibinfo  {journal} {Phys. Rev. C}\ }\textbf {\bibinfo
  {volume} {19}},\ \bibinfo {pages} {164} (\bibinfo {year} {1979})}\BibitemShut
  {NoStop}%
\bibitem [{\citenamefont {Warburton}\ \emph {et~al.}(1990)\citenamefont
  {Warburton}, \citenamefont {Becker},\ and\ \citenamefont {Brown}}]{WAR91}%
  \BibitemOpen
  \bibfield  {author} {\bibinfo {author} {\bibfnamefont {E.~K.}\ \bibnamefont
  {Warburton}}, \bibinfo {author} {\bibfnamefont {J.~A.}\ \bibnamefont
  {Becker}}, \ and\ \bibinfo {author} {\bibfnamefont {B.~A.}\ \bibnamefont
  {Brown}},\ }\href {\doibase 10.1103/PhysRevC.41.1147} {\bibfield  {journal}
  {\bibinfo  {journal} {Phys. Rev. C}\ }\textbf {\bibinfo {volume} {41}},\
  \bibinfo {pages} {1147} (\bibinfo {year} {1990})}\BibitemShut {NoStop}%
\bibitem [{\citenamefont {Sorlin}\ and\ \citenamefont {Porquet}(2008)}]{SOR08}%
  \BibitemOpen
  \bibfield  {author} {\bibinfo {author} {\bibfnamefont {O.}~\bibnamefont
  {Sorlin}}\ and\ \bibinfo {author} {\bibfnamefont {M.-G.}\ \bibnamefont
  {Porquet}},\ }\href {\doibase 10.1016/j.ppnp.2008.05.001} {\bibfield
  {journal} {\bibinfo  {journal} {Progress in Particle and Nuclear Physics}\
  }\textbf {\bibinfo {volume} {61}},\ \bibinfo {pages} {602} (\bibinfo {year}
  {2008})}\BibitemShut {NoStop}%
\bibitem [{\citenamefont {Nakamura}\ \emph {et~al.}(2009)\citenamefont
  {Nakamura}, \citenamefont {Kobayashi}, \citenamefont {Kondo}, \citenamefont
  {Satou}, \citenamefont {Aoi}, \citenamefont {Baba}, \citenamefont {Deguchi},
  \citenamefont {Fukuda}, \citenamefont {Gibelin}, \citenamefont {Inabe},
  \citenamefont {Ishihara}, \citenamefont {Kameda}, \citenamefont {Kawada},
  \citenamefont {Kubo}, \citenamefont {Kusaka}, \citenamefont {Mengoni},
  \citenamefont {Motobayashi}, \citenamefont {Ohnishi}, \citenamefont {Ohtake},
  \citenamefont {Orr}, \citenamefont {Otsu}, \citenamefont {Otsuka},
  \citenamefont {Saito}, \citenamefont {Sakurai}, \citenamefont {Shimoura},
  \citenamefont {Sumikama}, \citenamefont {Takeda}, \citenamefont {Takeshita},
  \citenamefont {Takechi}, \citenamefont {Takeuchi}, \citenamefont {Tanaka},
  \citenamefont {Tanaka}, \citenamefont {Tanaka}, \citenamefont {Togano},
  \citenamefont {Utsuno}, \citenamefont {Yoneda}, \citenamefont {Yoshida},\
  and\ \citenamefont {Yoshida}}]{NAKA09}%
  \BibitemOpen
  \bibfield  {author} {\bibinfo {author} {\bibfnamefont {T.}~\bibnamefont
  {Nakamura}}, \bibinfo {author} {\bibfnamefont {N.}~\bibnamefont {Kobayashi}},
  \bibinfo {author} {\bibfnamefont {Y.}~\bibnamefont {Kondo}}, \bibinfo
  {author} {\bibfnamefont {Y.}~\bibnamefont {Satou}}, \bibinfo {author}
  {\bibfnamefont {N.}~\bibnamefont {Aoi}}, \bibinfo {author} {\bibfnamefont
  {H.}~\bibnamefont {Baba}}, \bibinfo {author} {\bibfnamefont {S.}~\bibnamefont
  {Deguchi}}, \bibinfo {author} {\bibfnamefont {N.}~\bibnamefont {Fukuda}},
  \bibinfo {author} {\bibfnamefont {J.}~\bibnamefont {Gibelin}}, \bibinfo
  {author} {\bibfnamefont {N.}~\bibnamefont {Inabe}}, \bibinfo {author}
  {\bibfnamefont {M.}~\bibnamefont {Ishihara}}, \bibinfo {author}
  {\bibfnamefont {D.}~\bibnamefont {Kameda}}, \bibinfo {author} {\bibfnamefont
  {Y.}~\bibnamefont {Kawada}}, \bibinfo {author} {\bibfnamefont
  {T.}~\bibnamefont {Kubo}}, \bibinfo {author} {\bibfnamefont {K.}~\bibnamefont
  {Kusaka}}, \bibinfo {author} {\bibfnamefont {A.}~\bibnamefont {Mengoni}},
  \bibinfo {author} {\bibfnamefont {T.}~\bibnamefont {Motobayashi}}, \bibinfo
  {author} {\bibfnamefont {T.}~\bibnamefont {Ohnishi}}, \bibinfo {author}
  {\bibfnamefont {M.}~\bibnamefont {Ohtake}}, \bibinfo {author} {\bibfnamefont
  {N.~A.}\ \bibnamefont {Orr}}, \bibinfo {author} {\bibfnamefont
  {H.}~\bibnamefont {Otsu}}, \bibinfo {author} {\bibfnamefont {T.}~\bibnamefont
  {Otsuka}}, \bibinfo {author} {\bibfnamefont {A.}~\bibnamefont {Saito}},
  \bibinfo {author} {\bibfnamefont {H.}~\bibnamefont {Sakurai}}, \bibinfo
  {author} {\bibfnamefont {S.}~\bibnamefont {Shimoura}}, \bibinfo {author}
  {\bibfnamefont {T.}~\bibnamefont {Sumikama}}, \bibinfo {author}
  {\bibfnamefont {H.}~\bibnamefont {Takeda}}, \bibinfo {author} {\bibfnamefont
  {E.}~\bibnamefont {Takeshita}}, \bibinfo {author} {\bibfnamefont
  {M.}~\bibnamefont {Takechi}}, \bibinfo {author} {\bibfnamefont
  {S.}~\bibnamefont {Takeuchi}}, \bibinfo {author} {\bibfnamefont
  {K.}~\bibnamefont {Tanaka}}, \bibinfo {author} {\bibfnamefont {K.~N.}\
  \bibnamefont {Tanaka}}, \bibinfo {author} {\bibfnamefont {N.}~\bibnamefont
  {Tanaka}}, \bibinfo {author} {\bibfnamefont {Y.}~\bibnamefont {Togano}},
  \bibinfo {author} {\bibfnamefont {Y.}~\bibnamefont {Utsuno}}, \bibinfo
  {author} {\bibfnamefont {K.}~\bibnamefont {Yoneda}}, \bibinfo {author}
  {\bibfnamefont {A.}~\bibnamefont {Yoshida}}, \ and\ \bibinfo {author}
  {\bibfnamefont {K.}~\bibnamefont {Yoshida}},\ }\href {\doibase
  10.1103/PhysRevLett.103.262501} {\bibfield  {journal} {\bibinfo  {journal}
  {Phys. Rev. Lett.}\ }\textbf {\bibinfo {volume} {103}},\ \bibinfo {pages}
  {262501} (\bibinfo {year} {2009})}\BibitemShut {NoStop}%
\bibitem [{\citenamefont {Motobayashi}\ \emph {et~al.}(1995)\citenamefont
  {Motobayashi}, \citenamefont {Ikeda}, \citenamefont {Ieki}, \citenamefont
  {Inoue}, \citenamefont {Iwasa}, \citenamefont {Kikuchi}, \citenamefont
  {Kurokawa}, \citenamefont {Moriya}, \citenamefont {Ogawa}, \citenamefont
  {Murakami}, \citenamefont {Shimoura}, \citenamefont {Yanagisawa},
  \citenamefont {Nakamura}, \citenamefont {Watanabe}, \citenamefont {Ishihara},
  \citenamefont {Teranishi}, \citenamefont {Okuno},\ and\ \citenamefont
  {Casten}}]{MOTO95}%
  \BibitemOpen
  \bibfield  {author} {\bibinfo {author} {\bibfnamefont {T.}~\bibnamefont
  {Motobayashi}}, \bibinfo {author} {\bibfnamefont {Y.}~\bibnamefont {Ikeda}},
  \bibinfo {author} {\bibfnamefont {K.}~\bibnamefont {Ieki}}, \bibinfo {author}
  {\bibfnamefont {M.}~\bibnamefont {Inoue}}, \bibinfo {author} {\bibfnamefont
  {N.}~\bibnamefont {Iwasa}}, \bibinfo {author} {\bibfnamefont
  {T.}~\bibnamefont {Kikuchi}}, \bibinfo {author} {\bibfnamefont
  {M.}~\bibnamefont {Kurokawa}}, \bibinfo {author} {\bibfnamefont
  {S.}~\bibnamefont {Moriya}}, \bibinfo {author} {\bibfnamefont
  {S.}~\bibnamefont {Ogawa}}, \bibinfo {author} {\bibfnamefont
  {H.}~\bibnamefont {Murakami}}, \bibinfo {author} {\bibfnamefont
  {S.}~\bibnamefont {Shimoura}}, \bibinfo {author} {\bibfnamefont
  {Y.}~\bibnamefont {Yanagisawa}}, \bibinfo {author} {\bibfnamefont
  {T.}~\bibnamefont {Nakamura}}, \bibinfo {author} {\bibfnamefont
  {Y.}~\bibnamefont {Watanabe}}, \bibinfo {author} {\bibfnamefont
  {M.}~\bibnamefont {Ishihara}}, \bibinfo {author} {\bibfnamefont
  {T.}~\bibnamefont {Teranishi}}, \bibinfo {author} {\bibfnamefont
  {H.}~\bibnamefont {Okuno}}, \ and\ \bibinfo {author} {\bibfnamefont
  {R.}~\bibnamefont {Casten}},\ }\href {\doibase
  https://doi.org/10.1016/0370-2693(95)00012-A} {\bibfield  {journal} {\bibinfo
   {journal} {Phys. Lett. B}\ }\textbf {\bibinfo {volume} {346}},\ \bibinfo
  {pages} {9 } (\bibinfo {year} {1995})}\BibitemShut {NoStop}%
\bibitem [{\citenamefont {Utsuno}\ \emph {et~al.}(2004)\citenamefont {Utsuno},
  \citenamefont {Otsuka}, \citenamefont {Glasmacher}, \citenamefont
  {Mizusaki},\ and\ \citenamefont {Honma}}]{Utsuno04}%
  \BibitemOpen
  \bibfield  {author} {\bibinfo {author} {\bibfnamefont {Y.}~\bibnamefont
  {Utsuno}}, \bibinfo {author} {\bibfnamefont {T.}~\bibnamefont {Otsuka}},
  \bibinfo {author} {\bibfnamefont {T.}~\bibnamefont {Glasmacher}}, \bibinfo
  {author} {\bibfnamefont {T.}~\bibnamefont {Mizusaki}}, \ and\ \bibinfo
  {author} {\bibfnamefont {M.}~\bibnamefont {Honma}},\ }\href {\doibase
  10.1103/PhysRevC.70.044307} {\bibfield  {journal} {\bibinfo  {journal} {Phys.
  Rev. C}\ }\textbf {\bibinfo {volume} {70}},\ \bibinfo {pages} {044307}
  (\bibinfo {year} {2004})}\BibitemShut {NoStop}%
\bibitem [{\citenamefont {Poves}\ and\ \citenamefont
  {Retamosa}(1987)}]{Poves87}%
  \BibitemOpen
  \bibfield  {author} {\bibinfo {author} {\bibfnamefont {A.}~\bibnamefont
  {Poves}}\ and\ \bibinfo {author} {\bibfnamefont {J.}~\bibnamefont
  {Retamosa}},\ }\href {\doibase https://doi.org/10.1016/0370-2693(87)90171-7}
  {\bibfield  {journal} {\bibinfo  {journal} {Phys. Lett. B}\ }\textbf
  {\bibinfo {volume} {184}},\ \bibinfo {pages} {311 } (\bibinfo {year}
  {1987})}\BibitemShut {NoStop}%
\bibitem [{\citenamefont {{E. Caurier}}\ \emph {et~al.}(2002)\citenamefont {{E.
  Caurier}}, \citenamefont {{F. Nowacki}},\ and\ \citenamefont {{A.
  Poves}}}]{Poves02}%
  \BibitemOpen
  \bibfield  {author} {\bibinfo {author} {\bibnamefont {{E. Caurier}}},
  \bibinfo {author} {\bibnamefont {{F. Nowacki}}}, \ and\ \bibinfo {author}
  {\bibnamefont {{A. Poves}}},\ }\href {\doibase 10.1140/epja/i2001-10243-7}
  {\bibfield  {journal} {\bibinfo  {journal} {Eur. Phys. J. A}\ }\textbf
  {\bibinfo {volume} {15}},\ \bibinfo {pages} {145} (\bibinfo {year}
  {2002})}\BibitemShut {NoStop}%
\bibitem [{\citenamefont {Brown}\ and\ \citenamefont
  {Richter}(2006)}]{Brown06}%
  \BibitemOpen
  \bibfield  {author} {\bibinfo {author} {\bibfnamefont {B.~A.}\ \bibnamefont
  {Brown}}\ and\ \bibinfo {author} {\bibfnamefont {W.~A.}\ \bibnamefont
  {Richter}},\ }\href {\doibase 10.1103/PhysRevC.74.034315} {\bibfield
  {journal} {\bibinfo  {journal} {Phys. Rev. C}\ }\textbf {\bibinfo {volume}
  {74}},\ \bibinfo {pages} {034315} (\bibinfo {year} {2006})}\BibitemShut
  {NoStop}%
\bibitem [{\citenamefont {Ahn}\ \emph {et~al.}(2019)\citenamefont {Ahn},
  \citenamefont {Fukuda}, \citenamefont {Geissel}, \citenamefont {Inabe},
  \citenamefont {Iwasa}, \citenamefont {Kubo}, \citenamefont {Kusaka},
  \citenamefont {Morrissey}, \citenamefont {Murai}, \citenamefont {Nakamura},
  \citenamefont {Ohtake}, \citenamefont {Otsu}, \citenamefont {Sato},
  \citenamefont {Sherrill}, \citenamefont {Shimizu}, \citenamefont {Suzuki},
  \citenamefont {Takeda}, \citenamefont {Tarasov}, \citenamefont {Ueno},
  \citenamefont {Yanagisawa},\ and\ \citenamefont {Yoshida}}]{Ahn19}%
  \BibitemOpen
  \bibfield  {author} {\bibinfo {author} {\bibfnamefont {D.~S.}\ \bibnamefont
  {Ahn}}, \bibinfo {author} {\bibfnamefont {N.}~\bibnamefont {Fukuda}},
  \bibinfo {author} {\bibfnamefont {H.}~\bibnamefont {Geissel}}, \bibinfo
  {author} {\bibfnamefont {N.}~\bibnamefont {Inabe}}, \bibinfo {author}
  {\bibfnamefont {N.}~\bibnamefont {Iwasa}}, \bibinfo {author} {\bibfnamefont
  {T.}~\bibnamefont {Kubo}}, \bibinfo {author} {\bibfnamefont {K.}~\bibnamefont
  {Kusaka}}, \bibinfo {author} {\bibfnamefont {D.~J.}\ \bibnamefont
  {Morrissey}}, \bibinfo {author} {\bibfnamefont {D.}~\bibnamefont {Murai}},
  \bibinfo {author} {\bibfnamefont {T.}~\bibnamefont {Nakamura}}, \bibinfo
  {author} {\bibfnamefont {M.}~\bibnamefont {Ohtake}}, \bibinfo {author}
  {\bibfnamefont {H.}~\bibnamefont {Otsu}}, \bibinfo {author} {\bibfnamefont
  {H.}~\bibnamefont {Sato}}, \bibinfo {author} {\bibfnamefont {B.~M.}\
  \bibnamefont {Sherrill}}, \bibinfo {author} {\bibfnamefont {Y.}~\bibnamefont
  {Shimizu}}, \bibinfo {author} {\bibfnamefont {H.}~\bibnamefont {Suzuki}},
  \bibinfo {author} {\bibfnamefont {H.}~\bibnamefont {Takeda}}, \bibinfo
  {author} {\bibfnamefont {O.~B.}\ \bibnamefont {Tarasov}}, \bibinfo {author}
  {\bibfnamefont {H.}~\bibnamefont {Ueno}}, \bibinfo {author} {\bibfnamefont
  {Y.}~\bibnamefont {Yanagisawa}}, \ and\ \bibinfo {author} {\bibfnamefont
  {K.}~\bibnamefont {Yoshida}},\ }\href {\doibase
  10.1103/PhysRevLett.123.212501} {\bibfield  {journal} {\bibinfo  {journal}
  {Phys. Rev. Lett.}\ }\textbf {\bibinfo {volume} {123}},\ \bibinfo {pages}
  {212501} (\bibinfo {year} {2019})}\BibitemShut {NoStop}%
\bibitem [{\citenamefont {Elekes}\ \emph {et~al.}(2004)\citenamefont {Elekes},
  \citenamefont {Dombrádi}, \citenamefont {Saito}, \citenamefont {Aoi},
  \citenamefont {Baba}, \citenamefont {Demichi}, \citenamefont {Fülöp},
  \citenamefont {Gibelin}, \citenamefont {Gomi}, \citenamefont {Hasegawa},
  \citenamefont {Imai}, \citenamefont {Ishihara}, \citenamefont {Iwasaki},
  \citenamefont {Kanno}, \citenamefont {Kawai}, \citenamefont {Kishida},
  \citenamefont {Kubo}, \citenamefont {Kurita}, \citenamefont {Matsuyama},
  \citenamefont {Michimasa}, \citenamefont {Minemura}, \citenamefont
  {Motobayashi}, \citenamefont {Notani}, \citenamefont {Ohnishi}, \citenamefont
  {Ong}, \citenamefont {Ota}, \citenamefont {Ozawa}, \citenamefont {Sakai},
  \citenamefont {Sakurai}, \citenamefont {Shimoura}, \citenamefont {Takeshita},
  \citenamefont {Takeuchi}, \citenamefont {Tamaki}, \citenamefont {Togano},
  \citenamefont {Yamada}, \citenamefont {Yanagisawa},\ and\ \citenamefont
  {Yoneda}}]{Elekes04}%
  \BibitemOpen
  \bibfield  {author} {\bibinfo {author} {\bibfnamefont {Z.}~\bibnamefont
  {Elekes}}, \bibinfo {author} {\bibfnamefont {Z.}~\bibnamefont {Dombrádi}},
  \bibinfo {author} {\bibfnamefont {A.}~\bibnamefont {Saito}}, \bibinfo
  {author} {\bibfnamefont {N.}~\bibnamefont {Aoi}}, \bibinfo {author}
  {\bibfnamefont {H.}~\bibnamefont {Baba}}, \bibinfo {author} {\bibfnamefont
  {K.}~\bibnamefont {Demichi}}, \bibinfo {author} {\bibfnamefont
  {Z.}~\bibnamefont {Fülöp}}, \bibinfo {author} {\bibfnamefont
  {J.}~\bibnamefont {Gibelin}}, \bibinfo {author} {\bibfnamefont
  {T.}~\bibnamefont {Gomi}}, \bibinfo {author} {\bibfnamefont {H.}~\bibnamefont
  {Hasegawa}}, \bibinfo {author} {\bibfnamefont {N.}~\bibnamefont {Imai}},
  \bibinfo {author} {\bibfnamefont {M.}~\bibnamefont {Ishihara}}, \bibinfo
  {author} {\bibfnamefont {H.}~\bibnamefont {Iwasaki}}, \bibinfo {author}
  {\bibfnamefont {S.}~\bibnamefont {Kanno}}, \bibinfo {author} {\bibfnamefont
  {S.}~\bibnamefont {Kawai}}, \bibinfo {author} {\bibfnamefont
  {T.}~\bibnamefont {Kishida}}, \bibinfo {author} {\bibfnamefont
  {T.}~\bibnamefont {Kubo}}, \bibinfo {author} {\bibfnamefont {K.}~\bibnamefont
  {Kurita}}, \bibinfo {author} {\bibfnamefont {Y.}~\bibnamefont {Matsuyama}},
  \bibinfo {author} {\bibfnamefont {S.}~\bibnamefont {Michimasa}}, \bibinfo
  {author} {\bibfnamefont {T.}~\bibnamefont {Minemura}}, \bibinfo {author}
  {\bibfnamefont {T.}~\bibnamefont {Motobayashi}}, \bibinfo {author}
  {\bibfnamefont {M.}~\bibnamefont {Notani}}, \bibinfo {author} {\bibfnamefont
  {T.}~\bibnamefont {Ohnishi}}, \bibinfo {author} {\bibfnamefont
  {H.}~\bibnamefont {Ong}}, \bibinfo {author} {\bibfnamefont {S.}~\bibnamefont
  {Ota}}, \bibinfo {author} {\bibfnamefont {A.}~\bibnamefont {Ozawa}}, \bibinfo
  {author} {\bibfnamefont {H.}~\bibnamefont {Sakai}}, \bibinfo {author}
  {\bibfnamefont {H.}~\bibnamefont {Sakurai}}, \bibinfo {author} {\bibfnamefont
  {S.}~\bibnamefont {Shimoura}}, \bibinfo {author} {\bibfnamefont
  {E.}~\bibnamefont {Takeshita}}, \bibinfo {author} {\bibfnamefont
  {S.}~\bibnamefont {Takeuchi}}, \bibinfo {author} {\bibfnamefont
  {M.}~\bibnamefont {Tamaki}}, \bibinfo {author} {\bibfnamefont
  {Y.}~\bibnamefont {Togano}}, \bibinfo {author} {\bibfnamefont
  {K.}~\bibnamefont {Yamada}}, \bibinfo {author} {\bibfnamefont
  {Y.}~\bibnamefont {Yanagisawa}}, \ and\ \bibinfo {author} {\bibfnamefont
  {K.}~\bibnamefont {Yoneda}},\ }\href {\doibase
  https://doi.org/10.1016/j.physletb.2004.08.028} {\bibfield  {journal}
  {\bibinfo  {journal} {Phys. Lett. B}\ }\textbf {\bibinfo {volume} {599}},\
  \bibinfo {pages} {17 } (\bibinfo {year} {2004})}\BibitemShut {NoStop}%
\bibitem [{\citenamefont {Jurado}\ \emph {et~al.}(2007)\citenamefont {Jurado},
  \citenamefont {Savajols}, \citenamefont {Mittig}, \citenamefont {Orr},
  \citenamefont {Roussel-Chomaz}, \citenamefont {Baiborodin}, \citenamefont
  {Catford}, \citenamefont {Chartier}, \citenamefont {Demonchy}, \citenamefont
  {Dlouhý}, \citenamefont {Gillibert}, \citenamefont {Giot}, \citenamefont
  {Khouaja}, \citenamefont {Lépine-Szily}, \citenamefont {Lukyanov},
  \citenamefont {Mrazek}, \citenamefont {Penionzhkevich}, \citenamefont {Pita},
  \citenamefont {Rousseau},\ and\ \citenamefont {Villari}}]{Jurado07}%
  \BibitemOpen
  \bibfield  {author} {\bibinfo {author} {\bibfnamefont {B.}~\bibnamefont
  {Jurado}}, \bibinfo {author} {\bibfnamefont {H.}~\bibnamefont {Savajols}},
  \bibinfo {author} {\bibfnamefont {W.}~\bibnamefont {Mittig}}, \bibinfo
  {author} {\bibfnamefont {N.}~\bibnamefont {Orr}}, \bibinfo {author}
  {\bibfnamefont {P.}~\bibnamefont {Roussel-Chomaz}}, \bibinfo {author}
  {\bibfnamefont {D.}~\bibnamefont {Baiborodin}}, \bibinfo {author}
  {\bibfnamefont {W.}~\bibnamefont {Catford}}, \bibinfo {author} {\bibfnamefont
  {M.}~\bibnamefont {Chartier}}, \bibinfo {author} {\bibfnamefont
  {C.}~\bibnamefont {Demonchy}}, \bibinfo {author} {\bibfnamefont
  {Z.}~\bibnamefont {Dlouhý}}, \bibinfo {author} {\bibfnamefont
  {A.}~\bibnamefont {Gillibert}}, \bibinfo {author} {\bibfnamefont
  {L.}~\bibnamefont {Giot}}, \bibinfo {author} {\bibfnamefont {A.}~\bibnamefont
  {Khouaja}}, \bibinfo {author} {\bibfnamefont {A.}~\bibnamefont
  {Lépine-Szily}}, \bibinfo {author} {\bibfnamefont {S.}~\bibnamefont
  {Lukyanov}}, \bibinfo {author} {\bibfnamefont {J.}~\bibnamefont {Mrazek}},
  \bibinfo {author} {\bibfnamefont {Y.}~\bibnamefont {Penionzhkevich}},
  \bibinfo {author} {\bibfnamefont {S.}~\bibnamefont {Pita}}, \bibinfo {author}
  {\bibfnamefont {M.}~\bibnamefont {Rousseau}}, \ and\ \bibinfo {author}
  {\bibfnamefont {A.}~\bibnamefont {Villari}},\ }\href {\doibase
  https://doi.org/10.1016/j.physletb.2007.04.006} {\bibfield  {journal}
  {\bibinfo  {journal} {Phys. Lett. B}\ }\textbf {\bibinfo {volume} {649}},\
  \bibinfo {pages} {43 } (\bibinfo {year} {2007})}\BibitemShut {NoStop}%
\bibitem [{\citenamefont {Christian}\ \emph
  {et~al.}(2012{\natexlab{a}})\citenamefont {Christian}, \citenamefont {Frank},
  \citenamefont {Ash}, \citenamefont {Baumann}, \citenamefont {Bazin},
  \citenamefont {Brown}, \citenamefont {DeYoung}, \citenamefont {Finck},
  \citenamefont {Gade}, \citenamefont {Grinyer}, \citenamefont {Grovom},
  \citenamefont {Hinnefeld}, \citenamefont {Lunderberg}, \citenamefont
  {Luther}, \citenamefont {Mosby}, \citenamefont {Mosby}, \citenamefont {Nagi},
  \citenamefont {Peaslee}, \citenamefont {Rogers}, \citenamefont {Smith},
  \citenamefont {Snyder}, \citenamefont {Spyrou}, \citenamefont {Strongman},
  \citenamefont {Thoennessen}, \citenamefont {Warren}, \citenamefont
  {Weisshaar},\ and\ \citenamefont {Wersal}}]{CHRIS1}%
  \BibitemOpen
  \bibfield  {author} {\bibinfo {author} {\bibfnamefont {G.}~\bibnamefont
  {Christian}}, \bibinfo {author} {\bibfnamefont {N.}~\bibnamefont {Frank}},
  \bibinfo {author} {\bibfnamefont {S.}~\bibnamefont {Ash}}, \bibinfo {author}
  {\bibfnamefont {T.}~\bibnamefont {Baumann}}, \bibinfo {author} {\bibfnamefont
  {D.}~\bibnamefont {Bazin}}, \bibinfo {author} {\bibfnamefont
  {J.}~\bibnamefont {Brown}}, \bibinfo {author} {\bibfnamefont {P.~A.}\
  \bibnamefont {DeYoung}}, \bibinfo {author} {\bibfnamefont {J.~E.}\
  \bibnamefont {Finck}}, \bibinfo {author} {\bibfnamefont {A.}~\bibnamefont
  {Gade}}, \bibinfo {author} {\bibfnamefont {G.~F.}\ \bibnamefont {Grinyer}},
  \bibinfo {author} {\bibfnamefont {A.}~\bibnamefont {Grovom}}, \bibinfo
  {author} {\bibfnamefont {J.~D.}\ \bibnamefont {Hinnefeld}}, \bibinfo {author}
  {\bibfnamefont {E.~M.}\ \bibnamefont {Lunderberg}}, \bibinfo {author}
  {\bibfnamefont {B.}~\bibnamefont {Luther}}, \bibinfo {author} {\bibfnamefont
  {M.}~\bibnamefont {Mosby}}, \bibinfo {author} {\bibfnamefont
  {S.}~\bibnamefont {Mosby}}, \bibinfo {author} {\bibfnamefont
  {T.}~\bibnamefont {Nagi}}, \bibinfo {author} {\bibfnamefont {G.~F.}\
  \bibnamefont {Peaslee}}, \bibinfo {author} {\bibfnamefont {W.~F.}\
  \bibnamefont {Rogers}}, \bibinfo {author} {\bibfnamefont {J.~K.}\
  \bibnamefont {Smith}}, \bibinfo {author} {\bibfnamefont {J.}~\bibnamefont
  {Snyder}}, \bibinfo {author} {\bibfnamefont {A.}~\bibnamefont {Spyrou}},
  \bibinfo {author} {\bibfnamefont {M.~J.}\ \bibnamefont {Strongman}}, \bibinfo
  {author} {\bibfnamefont {M.}~\bibnamefont {Thoennessen}}, \bibinfo {author}
  {\bibfnamefont {M.}~\bibnamefont {Warren}}, \bibinfo {author} {\bibfnamefont
  {D.}~\bibnamefont {Weisshaar}}, \ and\ \bibinfo {author} {\bibfnamefont
  {A.}~\bibnamefont {Wersal}},\ }\href {\doibase
  10.1103/PhysRevLett.108.032501} {\bibfield  {journal} {\bibinfo  {journal}
  {Phys. Rev. Lett.}\ }\textbf {\bibinfo {volume} {108}},\ \bibinfo {pages}
  {032501} (\bibinfo {year} {2012}{\natexlab{a}})}\BibitemShut {NoStop}%
\bibitem [{\citenamefont {Christian}\ \emph
  {et~al.}(2012{\natexlab{b}})\citenamefont {Christian}, \citenamefont {Frank},
  \citenamefont {Ash}, \citenamefont {Baumann}, \citenamefont {DeYoung},
  \citenamefont {Finck}, \citenamefont {Gade}, \citenamefont {Grinyer},
  \citenamefont {Luther}, \citenamefont {Mosby}, \citenamefont {Mosby},
  \citenamefont {Smith}, \citenamefont {Snyder}, \citenamefont {Spyrou},
  \citenamefont {Strongman}, \citenamefont {Thoennessen}, \citenamefont
  {Warren}, \citenamefont {Weisshaar},\ and\ \citenamefont {Wersal}}]{CHRIS2}%
  \BibitemOpen
  \bibfield  {author} {\bibinfo {author} {\bibfnamefont {G.}~\bibnamefont
  {Christian}}, \bibinfo {author} {\bibfnamefont {N.}~\bibnamefont {Frank}},
  \bibinfo {author} {\bibfnamefont {S.}~\bibnamefont {Ash}}, \bibinfo {author}
  {\bibfnamefont {T.}~\bibnamefont {Baumann}}, \bibinfo {author} {\bibfnamefont
  {P.~A.}\ \bibnamefont {DeYoung}}, \bibinfo {author} {\bibfnamefont {J.~E.}\
  \bibnamefont {Finck}}, \bibinfo {author} {\bibfnamefont {A.}~\bibnamefont
  {Gade}}, \bibinfo {author} {\bibfnamefont {G.~F.}\ \bibnamefont {Grinyer}},
  \bibinfo {author} {\bibfnamefont {B.}~\bibnamefont {Luther}}, \bibinfo
  {author} {\bibfnamefont {M.}~\bibnamefont {Mosby}}, \bibinfo {author}
  {\bibfnamefont {S.}~\bibnamefont {Mosby}}, \bibinfo {author} {\bibfnamefont
  {J.~K.}\ \bibnamefont {Smith}}, \bibinfo {author} {\bibfnamefont
  {J.}~\bibnamefont {Snyder}}, \bibinfo {author} {\bibfnamefont
  {A.}~\bibnamefont {Spyrou}}, \bibinfo {author} {\bibfnamefont {M.~J.}\
  \bibnamefont {Strongman}}, \bibinfo {author} {\bibfnamefont {M.}~\bibnamefont
  {Thoennessen}}, \bibinfo {author} {\bibfnamefont {M.}~\bibnamefont {Warren}},
  \bibinfo {author} {\bibfnamefont {D.}~\bibnamefont {Weisshaar}}, \ and\
  \bibinfo {author} {\bibfnamefont {A.}~\bibnamefont {Wersal}},\ }\href
  {\doibase 10.1103/PhysRevC.85.034327} {\bibfield  {journal} {\bibinfo
  {journal} {Phys. Rev. C}\ }\textbf {\bibinfo {volume} {85}},\ \bibinfo
  {pages} {034327} (\bibinfo {year} {2012}{\natexlab{b}})}\BibitemShut
  {NoStop}%
\bibitem [{\citenamefont {Gaudefroy}\ \emph {et~al.}(2012)\citenamefont
  {Gaudefroy}, \citenamefont {Mittig}, \citenamefont {Orr}, \citenamefont
  {Varet}, \citenamefont {Chartier}, \citenamefont {Roussel-Chomaz},
  \citenamefont {Ebran}, \citenamefont {Fern\'andez-Dom\'{\i}nguez},
  \citenamefont {Fr\'emont}, \citenamefont {Gangnant}, \citenamefont
  {Gillibert}, \citenamefont {Gr\'evy}, \citenamefont {Libin}, \citenamefont
  {Maslov}, \citenamefont {Paschalis}, \citenamefont {Pietras}, \citenamefont
  {Penionzhkevich}, \citenamefont {Spitaels},\ and\ \citenamefont
  {Villari}}]{GAUDE}%
  \BibitemOpen
  \bibfield  {author} {\bibinfo {author} {\bibfnamefont {L.}~\bibnamefont
  {Gaudefroy}}, \bibinfo {author} {\bibfnamefont {W.}~\bibnamefont {Mittig}},
  \bibinfo {author} {\bibfnamefont {N.~A.}\ \bibnamefont {Orr}}, \bibinfo
  {author} {\bibfnamefont {S.}~\bibnamefont {Varet}}, \bibinfo {author}
  {\bibfnamefont {M.}~\bibnamefont {Chartier}}, \bibinfo {author}
  {\bibfnamefont {P.}~\bibnamefont {Roussel-Chomaz}}, \bibinfo {author}
  {\bibfnamefont {J.~P.}\ \bibnamefont {Ebran}}, \bibinfo {author}
  {\bibfnamefont {B.}~\bibnamefont {Fern\'andez-Dom\'{\i}nguez}}, \bibinfo
  {author} {\bibfnamefont {G.}~\bibnamefont {Fr\'emont}}, \bibinfo {author}
  {\bibfnamefont {P.}~\bibnamefont {Gangnant}}, \bibinfo {author}
  {\bibfnamefont {A.}~\bibnamefont {Gillibert}}, \bibinfo {author}
  {\bibfnamefont {S.}~\bibnamefont {Gr\'evy}}, \bibinfo {author} {\bibfnamefont
  {J.~F.}\ \bibnamefont {Libin}}, \bibinfo {author} {\bibfnamefont {V.~A.}\
  \bibnamefont {Maslov}}, \bibinfo {author} {\bibfnamefont {S.}~\bibnamefont
  {Paschalis}}, \bibinfo {author} {\bibfnamefont {B.}~\bibnamefont {Pietras}},
  \bibinfo {author} {\bibfnamefont {Y.-E.}\ \bibnamefont {Penionzhkevich}},
  \bibinfo {author} {\bibfnamefont {C.}~\bibnamefont {Spitaels}}, \ and\
  \bibinfo {author} {\bibfnamefont {A.~C.~C.}\ \bibnamefont {Villari}},\ }\href
  {\doibase 10.1103/PhysRevLett.109.202503} {\bibfield  {journal} {\bibinfo
  {journal} {Phys. Rev. Lett.}\ }\textbf {\bibinfo {volume} {109}},\ \bibinfo
  {pages} {202503} (\bibinfo {year} {2012})}\BibitemShut {NoStop}%
\bibitem [{\citenamefont {Wang}\ \emph {et~al.}(2017)\citenamefont {Wang},
  \citenamefont {Audi}, \citenamefont {Kondev}, \citenamefont {Huang},
  \citenamefont {Naimi},\ and\ \citenamefont {Xu}}]{WANG}%
  \BibitemOpen
  \bibfield  {author} {\bibinfo {author} {\bibfnamefont {M.}~\bibnamefont
  {Wang}}, \bibinfo {author} {\bibfnamefont {G.}~\bibnamefont {Audi}}, \bibinfo
  {author} {\bibfnamefont {F.~G.}\ \bibnamefont {Kondev}}, \bibinfo {author}
  {\bibfnamefont {W.}~\bibnamefont {Huang}}, \bibinfo {author} {\bibfnamefont
  {S.}~\bibnamefont {Naimi}}, \ and\ \bibinfo {author} {\bibfnamefont
  {X.}~\bibnamefont {Xu}},\ }\href {\doibase 10.1088/1674-1137/41/3/030003}
  {\bibfield  {journal} {\bibinfo  {journal} {Chinese Physics C}\ }\textbf
  {\bibinfo {volume} {41}},\ \bibinfo {pages} {030003} (\bibinfo {year}
  {2017})}\BibitemShut {NoStop}%
\bibitem [{\citenamefont {Doornenbal}\ \emph {et~al.}(2017)\citenamefont
  {Doornenbal}, \citenamefont {Scheit}, \citenamefont {Takeuchi}, \citenamefont
  {Utsuno}, \citenamefont {Aoi}, \citenamefont {Li}, \citenamefont
  {Matsushita}, \citenamefont {Steppenbeck}, \citenamefont {Wang},
  \citenamefont {Baba}, \citenamefont {Ideguchi}, \citenamefont {Kobayashi},
  \citenamefont {Kondo}, \citenamefont {Lee}, \citenamefont {Michimasa},
  \citenamefont {Motobayashi}, \citenamefont {Otsuka}, \citenamefont {Sakurai},
  \citenamefont {Takechi}, \citenamefont {Togano},\ and\ \citenamefont
  {Yoneda}}]{DOOR}%
  \BibitemOpen
  \bibfield  {author} {\bibinfo {author} {\bibfnamefont {P.}~\bibnamefont
  {Doornenbal}}, \bibinfo {author} {\bibfnamefont {H.}~\bibnamefont {Scheit}},
  \bibinfo {author} {\bibfnamefont {S.}~\bibnamefont {Takeuchi}}, \bibinfo
  {author} {\bibfnamefont {Y.}~\bibnamefont {Utsuno}}, \bibinfo {author}
  {\bibfnamefont {N.}~\bibnamefont {Aoi}}, \bibinfo {author} {\bibfnamefont
  {K.}~\bibnamefont {Li}}, \bibinfo {author} {\bibfnamefont {M.}~\bibnamefont
  {Matsushita}}, \bibinfo {author} {\bibfnamefont {D.}~\bibnamefont
  {Steppenbeck}}, \bibinfo {author} {\bibfnamefont {H.}~\bibnamefont {Wang}},
  \bibinfo {author} {\bibfnamefont {H.}~\bibnamefont {Baba}}, \bibinfo {author}
  {\bibfnamefont {E.}~\bibnamefont {Ideguchi}}, \bibinfo {author}
  {\bibfnamefont {N.}~\bibnamefont {Kobayashi}}, \bibinfo {author}
  {\bibfnamefont {Y.}~\bibnamefont {Kondo}}, \bibinfo {author} {\bibfnamefont
  {J.}~\bibnamefont {Lee}}, \bibinfo {author} {\bibfnamefont {S.}~\bibnamefont
  {Michimasa}}, \bibinfo {author} {\bibfnamefont {T.}~\bibnamefont
  {Motobayashi}}, \bibinfo {author} {\bibfnamefont {T.}~\bibnamefont {Otsuka}},
  \bibinfo {author} {\bibfnamefont {H.}~\bibnamefont {Sakurai}}, \bibinfo
  {author} {\bibfnamefont {M.}~\bibnamefont {Takechi}}, \bibinfo {author}
  {\bibfnamefont {Y.}~\bibnamefont {Togano}}, \ and\ \bibinfo {author}
  {\bibfnamefont {K.}~\bibnamefont {Yoneda}},\ }\href {\doibase
  10.1103/PhysRevC.95.041301} {\bibfield  {journal} {\bibinfo  {journal} {Phys.
  Rev. C}\ }\textbf {\bibinfo {volume} {95}},\ \bibinfo {pages} {041301}
  (\bibinfo {year} {2017})}\BibitemShut {NoStop}%
\bibitem [{\citenamefont {Utsuno}\ \emph {et~al.}(1999)\citenamefont {Utsuno},
  \citenamefont {Otsuka}, \citenamefont {Mizusaki},\ and\ \citenamefont
  {Honma}}]{Utsuno1999}%
  \BibitemOpen
  \bibfield  {author} {\bibinfo {author} {\bibfnamefont {Y.}~\bibnamefont
  {Utsuno}}, \bibinfo {author} {\bibfnamefont {T.}~\bibnamefont {Otsuka}},
  \bibinfo {author} {\bibfnamefont {T.}~\bibnamefont {Mizusaki}}, \ and\
  \bibinfo {author} {\bibfnamefont {M.}~\bibnamefont {Honma}},\ }\href
  {\doibase 10.1103/PhysRevC.60.054315} {\bibfield  {journal} {\bibinfo
  {journal} {Phys. Rev. C}\ }\textbf {\bibinfo {volume} {60}},\ \bibinfo
  {pages} {054315} (\bibinfo {year} {1999})}\BibitemShut {NoStop}%
\bibitem [{\citenamefont {Tanihata}\ \emph {et~al.}(2013)\citenamefont
  {Tanihata}, \citenamefont {Savajols},\ and\ \citenamefont
  {Kanungo}}]{Tanihata13}%
  \BibitemOpen
  \bibfield  {author} {\bibinfo {author} {\bibfnamefont {I.}~\bibnamefont
  {Tanihata}}, \bibinfo {author} {\bibfnamefont {H.}~\bibnamefont {Savajols}},
  \ and\ \bibinfo {author} {\bibfnamefont {R.}~\bibnamefont {Kanungo}},\ }\href
  {\doibase https://doi.org/10.1016/j.ppnp.2012.07.001} {\bibfield  {journal}
  {\bibinfo  {journal} {Progress in Particle and Nuclear Physics}\ }\textbf
  {\bibinfo {volume} {68}},\ \bibinfo {pages} {215 } (\bibinfo {year}
  {2013})}\BibitemShut {NoStop}%
\bibitem [{\citenamefont {Horiuchi}\ and\ \citenamefont
  {Suzuki}(2006)}]{Horiuchi06}%
  \BibitemOpen
  \bibfield  {author} {\bibinfo {author} {\bibfnamefont {W.}~\bibnamefont
  {Horiuchi}}\ and\ \bibinfo {author} {\bibfnamefont {Y.}~\bibnamefont
  {Suzuki}},\ }\href {\doibase 10.1103/PhysRevC.74.034311} {\bibfield
  {journal} {\bibinfo  {journal} {Phys. Rev. C}\ }\textbf {\bibinfo {volume}
  {74}},\ \bibinfo {pages} {034311} (\bibinfo {year} {2006})}\BibitemShut
  {NoStop}%
\bibitem [{\citenamefont {Togano}\ \emph {et~al.}(2016)\citenamefont {Togano},
  \citenamefont {Nakamura}, \citenamefont {Kondo}, \citenamefont {Tostevin},
  \citenamefont {Saito}, \citenamefont {Gibelin}, \citenamefont {Orr},
  \citenamefont {Achouri}, \citenamefont {Aumann}, \citenamefont {Baba},
  \citenamefont {Delaunay}, \citenamefont {Doornenbal}, \citenamefont {Fukuda},
  \citenamefont {Hwang}, \citenamefont {Inabe}, \citenamefont {Isobe},
  \citenamefont {Kameda}, \citenamefont {Kanno}, \citenamefont {Kim},
  \citenamefont {Kobayashi}, \citenamefont {Kobayashi}, \citenamefont {Kubo},
  \citenamefont {Leblond}, \citenamefont {Lee}, \citenamefont {Marqués},
  \citenamefont {Minakata}, \citenamefont {Motobayashi}, \citenamefont {Murai},
  \citenamefont {Murakami}, \citenamefont {Muto}, \citenamefont {Nakashima},
  \citenamefont {Nakatsuka}, \citenamefont {Navin}, \citenamefont {Nishi},
  \citenamefont {Ogoshi}, \citenamefont {Otsu}, \citenamefont {Sato},
  \citenamefont {Satou}, \citenamefont {Shimizu}, \citenamefont {Suzuki},
  \citenamefont {Takahashi}, \citenamefont {Takeda}, \citenamefont {Takeuchi},
  \citenamefont {Tanaka}, \citenamefont {Tuff}, \citenamefont {Vandebrouck},\
  and\ \citenamefont {Yoneda}}]{Togano16}%
  \BibitemOpen
  \bibfield  {author} {\bibinfo {author} {\bibfnamefont {Y.}~\bibnamefont
  {Togano}}, \bibinfo {author} {\bibfnamefont {T.}~\bibnamefont {Nakamura}},
  \bibinfo {author} {\bibfnamefont {Y.}~\bibnamefont {Kondo}}, \bibinfo
  {author} {\bibfnamefont {J.}~\bibnamefont {Tostevin}}, \bibinfo {author}
  {\bibfnamefont {A.}~\bibnamefont {Saito}}, \bibinfo {author} {\bibfnamefont
  {J.}~\bibnamefont {Gibelin}}, \bibinfo {author} {\bibfnamefont
  {N.}~\bibnamefont {Orr}}, \bibinfo {author} {\bibfnamefont {N.}~\bibnamefont
  {Achouri}}, \bibinfo {author} {\bibfnamefont {T.}~\bibnamefont {Aumann}},
  \bibinfo {author} {\bibfnamefont {H.}~\bibnamefont {Baba}}, \bibinfo {author}
  {\bibfnamefont {F.}~\bibnamefont {Delaunay}}, \bibinfo {author}
  {\bibfnamefont {P.}~\bibnamefont {Doornenbal}}, \bibinfo {author}
  {\bibfnamefont {N.}~\bibnamefont {Fukuda}}, \bibinfo {author} {\bibfnamefont
  {J.}~\bibnamefont {Hwang}}, \bibinfo {author} {\bibfnamefont
  {N.}~\bibnamefont {Inabe}}, \bibinfo {author} {\bibfnamefont
  {T.}~\bibnamefont {Isobe}}, \bibinfo {author} {\bibfnamefont
  {D.}~\bibnamefont {Kameda}}, \bibinfo {author} {\bibfnamefont
  {D.}~\bibnamefont {Kanno}}, \bibinfo {author} {\bibfnamefont
  {S.}~\bibnamefont {Kim}}, \bibinfo {author} {\bibfnamefont {N.}~\bibnamefont
  {Kobayashi}}, \bibinfo {author} {\bibfnamefont {T.}~\bibnamefont
  {Kobayashi}}, \bibinfo {author} {\bibfnamefont {T.}~\bibnamefont {Kubo}},
  \bibinfo {author} {\bibfnamefont {S.}~\bibnamefont {Leblond}}, \bibinfo
  {author} {\bibfnamefont {J.}~\bibnamefont {Lee}}, \bibinfo {author}
  {\bibfnamefont {F.}~\bibnamefont {Marqués}}, \bibinfo {author}
  {\bibfnamefont {R.}~\bibnamefont {Minakata}}, \bibinfo {author}
  {\bibfnamefont {T.}~\bibnamefont {Motobayashi}}, \bibinfo {author}
  {\bibfnamefont {D.}~\bibnamefont {Murai}}, \bibinfo {author} {\bibfnamefont
  {T.}~\bibnamefont {Murakami}}, \bibinfo {author} {\bibfnamefont
  {K.}~\bibnamefont {Muto}}, \bibinfo {author} {\bibfnamefont {T.}~\bibnamefont
  {Nakashima}}, \bibinfo {author} {\bibfnamefont {N.}~\bibnamefont
  {Nakatsuka}}, \bibinfo {author} {\bibfnamefont {A.}~\bibnamefont {Navin}},
  \bibinfo {author} {\bibfnamefont {S.}~\bibnamefont {Nishi}}, \bibinfo
  {author} {\bibfnamefont {S.}~\bibnamefont {Ogoshi}}, \bibinfo {author}
  {\bibfnamefont {H.}~\bibnamefont {Otsu}}, \bibinfo {author} {\bibfnamefont
  {H.}~\bibnamefont {Sato}}, \bibinfo {author} {\bibfnamefont {Y.}~\bibnamefont
  {Satou}}, \bibinfo {author} {\bibfnamefont {Y.}~\bibnamefont {Shimizu}},
  \bibinfo {author} {\bibfnamefont {H.}~\bibnamefont {Suzuki}}, \bibinfo
  {author} {\bibfnamefont {K.}~\bibnamefont {Takahashi}}, \bibinfo {author}
  {\bibfnamefont {H.}~\bibnamefont {Takeda}}, \bibinfo {author} {\bibfnamefont
  {S.}~\bibnamefont {Takeuchi}}, \bibinfo {author} {\bibfnamefont
  {R.}~\bibnamefont {Tanaka}}, \bibinfo {author} {\bibfnamefont
  {A.}~\bibnamefont {Tuff}}, \bibinfo {author} {\bibfnamefont {M.}~\bibnamefont
  {Vandebrouck}}, \ and\ \bibinfo {author} {\bibfnamefont {K.}~\bibnamefont
  {Yoneda}},\ }\href {\doibase https://doi.org/10.1016/j.physletb.2016.08.062}
  {\bibfield  {journal} {\bibinfo  {journal} {Phys. Lett. B}\ }\textbf
  {\bibinfo {volume} {761}},\ \bibinfo {pages} {412 } (\bibinfo {year}
  {2016})}\BibitemShut {NoStop}%
\bibitem [{\citenamefont {Zhukov}\ \emph {et~al.}(1993)\citenamefont {Zhukov},
  \citenamefont {Danilin}, \citenamefont {Fedorov}, \citenamefont {Bang},
  \citenamefont {Thompson},\ and\ \citenamefont {Vaagen}}]{Zhukov93}%
  \BibitemOpen
  \bibfield  {author} {\bibinfo {author} {\bibfnamefont {M.}~\bibnamefont
  {Zhukov}}, \bibinfo {author} {\bibfnamefont {B.}~\bibnamefont {Danilin}},
  \bibinfo {author} {\bibfnamefont {D.}~\bibnamefont {Fedorov}}, \bibinfo
  {author} {\bibfnamefont {J.}~\bibnamefont {Bang}}, \bibinfo {author}
  {\bibfnamefont {I.}~\bibnamefont {Thompson}}, \ and\ \bibinfo {author}
  {\bibfnamefont {J.}~\bibnamefont {Vaagen}},\ }\href {\doibase
  https://doi.org/10.1016/0370-1573(93)90141-Y} {\bibfield  {journal} {\bibinfo
   {journal} {Physics Reports}\ }\textbf {\bibinfo {volume} {231}},\ \bibinfo
  {pages} {151 } (\bibinfo {year} {1993})}\BibitemShut {NoStop}%
\bibitem [{\citenamefont {Nielsen}\ \emph {et~al.}(2001)\citenamefont
  {Nielsen}, \citenamefont {Fedorov}, \citenamefont {Jensen},\ and\
  \citenamefont {Garrido}}]{Nielsen01}%
  \BibitemOpen
  \bibfield  {author} {\bibinfo {author} {\bibfnamefont {E.}~\bibnamefont
  {Nielsen}}, \bibinfo {author} {\bibfnamefont {D.}~\bibnamefont {Fedorov}},
  \bibinfo {author} {\bibfnamefont {A.}~\bibnamefont {Jensen}}, \ and\ \bibinfo
  {author} {\bibfnamefont {E.}~\bibnamefont {Garrido}},\ }\href {\doibase
  https://doi.org/10.1016/S0370-1573(00)00107-1} {\bibfield  {journal}
  {\bibinfo  {journal} {Physics Reports}\ }\textbf {\bibinfo {volume} {347}},\
  \bibinfo {pages} {373 } (\bibinfo {year} {2001})}\BibitemShut {NoStop}%
\bibitem [{\citenamefont {Casal}(2018)}]{Casal18}%
  \BibitemOpen
  \bibfield  {author} {\bibinfo {author} {\bibfnamefont {J.}~\bibnamefont
  {Casal}},\ }\href {\doibase 10.1103/PhysRevC.97.034613} {\bibfield  {journal}
  {\bibinfo  {journal} {Phys. Rev. C}\ }\textbf {\bibinfo {volume} {97}},\
  \bibinfo {pages} {034613} (\bibinfo {year} {2018})}\BibitemShut {NoStop}%
\bibitem [{\citenamefont {Tolstikhin}\ \emph {et~al.}(1997)\citenamefont
  {Tolstikhin}, \citenamefont {Ostrovsky},\ and\ \citenamefont
  {Nakamura}}]{Tolstikhin97}%
  \BibitemOpen
  \bibfield  {author} {\bibinfo {author} {\bibfnamefont {O.~I.}\ \bibnamefont
  {Tolstikhin}}, \bibinfo {author} {\bibfnamefont {V.~N.}\ \bibnamefont
  {Ostrovsky}}, \ and\ \bibinfo {author} {\bibfnamefont {H.}~\bibnamefont
  {Nakamura}},\ }\href {\doibase 10.1103/PhysRevLett.79.2026} {\bibfield
  {journal} {\bibinfo  {journal} {Phys. Rev. Lett.}\ }\textbf {\bibinfo
  {volume} {79}},\ \bibinfo {pages} {2026} (\bibinfo {year}
  {1997})}\BibitemShut {NoStop}%
\bibitem [{\citenamefont {Descouvemont}\ \emph {et~al.}(2003)\citenamefont
  {Descouvemont}, \citenamefont {Daniel},\ and\ \citenamefont {Baye}}]{Desc03}%
  \BibitemOpen
  \bibfield  {author} {\bibinfo {author} {\bibfnamefont {P.}~\bibnamefont
  {Descouvemont}}, \bibinfo {author} {\bibfnamefont {C.}~\bibnamefont
  {Daniel}}, \ and\ \bibinfo {author} {\bibfnamefont {D.}~\bibnamefont
  {Baye}},\ }\href {\doibase 10.1103/PhysRevC.67.044309} {\bibfield  {journal}
  {\bibinfo  {journal} {Phys. Rev. C}\ }\textbf {\bibinfo {volume} {67}},\
  \bibinfo {pages} {044309} (\bibinfo {year} {2003})}\BibitemShut {NoStop}%
\bibitem [{\citenamefont {Matsumoto}\ \emph {et~al.}(2004)\citenamefont
  {Matsumoto}, \citenamefont {Hiyama}, \citenamefont {Yahiro}, \citenamefont
  {K.Ogata}, \citenamefont {Iseri},\ and\ \citenamefont
  {Kamimura}}]{Matsumoto04}%
  \BibitemOpen
  \bibfield  {author} {\bibinfo {author} {\bibfnamefont {T.}~\bibnamefont
  {Matsumoto}}, \bibinfo {author} {\bibfnamefont {E.}~\bibnamefont {Hiyama}},
  \bibinfo {author} {\bibfnamefont {M.}~\bibnamefont {Yahiro}}, \bibinfo
  {author} {\bibnamefont {K.Ogata}}, \bibinfo {author} {\bibfnamefont
  {Y.}~\bibnamefont {Iseri}}, \ and\ \bibinfo {author} {\bibfnamefont
  {M.}~\bibnamefont {Kamimura}},\ }\href {\doibase
  10.1016/j.nuclphysa.2004.04.089} {\bibfield  {journal} {\bibinfo  {journal}
  {Nucl. Phys. A}\ }\textbf {\bibinfo {volume} {738}},\ \bibinfo {pages} {471}
  (\bibinfo {year} {2004})}\BibitemShut {NoStop}%
\bibitem [{\citenamefont {Rodr\'{\i}guez-Gallardo}\ \emph
  {et~al.}(2005)\citenamefont {Rodr\'{\i}guez-Gallardo}, \citenamefont {Arias},
  \citenamefont {G\'omez-Camacho}, \citenamefont {Moro}, \citenamefont
  {Thompson},\ and\ \citenamefont {Tostevin}}]{MRoGa05}%
  \BibitemOpen
  \bibfield  {author} {\bibinfo {author} {\bibfnamefont {M.}~\bibnamefont
  {Rodr\'{\i}guez-Gallardo}}, \bibinfo {author} {\bibfnamefont {J.~M.}\
  \bibnamefont {Arias}}, \bibinfo {author} {\bibfnamefont {J.}~\bibnamefont
  {G\'omez-Camacho}}, \bibinfo {author} {\bibfnamefont {A.~M.}\ \bibnamefont
  {Moro}}, \bibinfo {author} {\bibfnamefont {I.~J.}\ \bibnamefont {Thompson}},
  \ and\ \bibinfo {author} {\bibfnamefont {J.~A.}\ \bibnamefont {Tostevin}},\
  }\href {\doibase 10.1103/PhysRevC.72.024007} {\bibfield  {journal} {\bibinfo
  {journal} {Phys. Rev. C}\ }\textbf {\bibinfo {volume} {72}},\ \bibinfo
  {pages} {024007} (\bibinfo {year} {2005})}\BibitemShut {NoStop}%
\bibitem [{\citenamefont {Casal}\ \emph {et~al.}(2013)\citenamefont {Casal},
  \citenamefont {Rodr\'{\i}guez-Gallardo},\ and\ \citenamefont
  {Arias}}]{Casal13}%
  \BibitemOpen
  \bibfield  {author} {\bibinfo {author} {\bibfnamefont {J.}~\bibnamefont
  {Casal}}, \bibinfo {author} {\bibfnamefont {M.}~\bibnamefont
  {Rodr\'{\i}guez-Gallardo}}, \ and\ \bibinfo {author} {\bibfnamefont {J.~M.}\
  \bibnamefont {Arias}},\ }\href {\doibase 10.1103/PhysRevC.88.014327}
  {\bibfield  {journal} {\bibinfo  {journal} {Phys. Rev. C}\ }\textbf {\bibinfo
  {volume} {88}},\ \bibinfo {pages} {014327} (\bibinfo {year}
  {2013})}\BibitemShut {NoStop}%
\bibitem [{\citenamefont {Casal}(2016)}]{CasalTh}%
  \BibitemOpen
  \bibfield  {author} {\bibinfo {author} {\bibfnamefont {J.}~\bibnamefont
  {Casal}},\ }\href {https://idus.us.es/xmlui/handle/11441/41814} {\enquote
  {\bibinfo {title} {Weakly-bound three-body nuclear systems: Structure,
  reactions and astrophysical implications},}\ }\bibinfo {howpublished} {Ph.D.
  thesis, Universidad de Sevilla} (\bibinfo {year} {2016})\BibitemShut
  {NoStop}%
\bibitem [{\citenamefont {Thompson}\ \emph {et~al.}(2004)\citenamefont
  {Thompson}, \citenamefont {Nunes},\ and\ \citenamefont
  {Danilin}}]{IJThompson04}%
  \BibitemOpen
  \bibfield  {author} {\bibinfo {author} {\bibfnamefont {I.}~\bibnamefont
  {Thompson}}, \bibinfo {author} {\bibfnamefont {F.}~\bibnamefont {Nunes}}, \
  and\ \bibinfo {author} {\bibfnamefont {B.}~\bibnamefont {Danilin}},\ }\href
  {\doibase https://doi.org/10.1016/j.cpc.2004.03.007} {\bibfield  {journal}
  {\bibinfo  {journal} {Computer Physics Communications}\ }\textbf {\bibinfo
  {volume} {161}},\ \bibinfo {pages} {87 } (\bibinfo {year}
  {2004})}\BibitemShut {NoStop}%
\bibitem [{\citenamefont {de~Diego}\ \emph {et~al.}(2010)\citenamefont
  {de~Diego}, \citenamefont {Garrido}, \citenamefont {Fedorov},\ and\
  \citenamefont {Jensen}}]{RdDiego10}%
  \BibitemOpen
  \bibfield  {author} {\bibinfo {author} {\bibfnamefont {R.}~\bibnamefont
  {de~Diego}}, \bibinfo {author} {\bibfnamefont {E.}~\bibnamefont {Garrido}},
  \bibinfo {author} {\bibfnamefont {D.~V.}\ \bibnamefont {Fedorov}}, \ and\
  \bibinfo {author} {\bibfnamefont {A.~S.}\ \bibnamefont {Jensen}},\ }\href
  {http://stacks.iop.org/0295-5075/90/i=5/a=52001} {\bibfield  {journal}
  {\bibinfo  {journal} {EPL (Europhys. Lett.)}\ }\textbf {\bibinfo {volume}
  {90}},\ \bibinfo {pages} {52001} (\bibinfo {year} {2010})}\BibitemShut
  {NoStop}%
\bibitem [{\citenamefont {Hagino}\ and\ \citenamefont {Sagawa}(2016)}]{HAG}%
  \BibitemOpen
  \bibfield  {author} {\bibinfo {author} {\bibfnamefont {K.}~\bibnamefont
  {Hagino}}\ and\ \bibinfo {author} {\bibfnamefont {H.}~\bibnamefont
  {Sagawa}},\ }\href {\doibase 10.1103/PhysRevC.93.034330} {\bibfield
  {journal} {\bibinfo  {journal} {Phys. Rev. C}\ }\textbf {\bibinfo {volume}
  {93}},\ \bibinfo {pages} {034330} (\bibinfo {year} {2016})}\BibitemShut
  {NoStop}%
\bibitem [{\citenamefont {Singh}\ \emph {et~al.}(2016)\citenamefont {Singh},
  \citenamefont {Fortunato}, \citenamefont {Vitturi},\ and\ \citenamefont
  {Chatterjee}}]{JS16}%
  \BibitemOpen
  \bibfield  {author} {\bibinfo {author} {\bibfnamefont {J.}~\bibnamefont
  {Singh}}, \bibinfo {author} {\bibfnamefont {L.}~\bibnamefont {Fortunato}},
  \bibinfo {author} {\bibfnamefont {A.}~\bibnamefont {Vitturi}}, \ and\
  \bibinfo {author} {\bibfnamefont {R.}~\bibnamefont {Chatterjee}},\ }\href
  {\doibase 10.1140/epja/i2016-16209-8} {\bibfield  {journal} {\bibinfo
  {journal} {Eur. Phys. J. A}\ }\textbf {\bibinfo {volume} {52}},\ \bibinfo
  {pages} {209} (\bibinfo {year} {2016})}\BibitemShut {NoStop}%
\bibitem [{\citenamefont {Lovell}\ \emph {et~al.}(2017)\citenamefont {Lovell},
  \citenamefont {Nunes},\ and\ \citenamefont {Thompson}}]{Lovell17}%
  \BibitemOpen
  \bibfield  {author} {\bibinfo {author} {\bibfnamefont {A.~E.}\ \bibnamefont
  {Lovell}}, \bibinfo {author} {\bibfnamefont {F.~M.}\ \bibnamefont {Nunes}}, \
  and\ \bibinfo {author} {\bibfnamefont {I.~J.}\ \bibnamefont {Thompson}},\
  }\href {\doibase 10.1103/PhysRevC.95.034605} {\bibfield  {journal} {\bibinfo
  {journal} {Phys. Rev. C}\ }\textbf {\bibinfo {volume} {95}},\ \bibinfo
  {pages} {034605} (\bibinfo {year} {2017})}\BibitemShut {NoStop}%
\bibitem [{\citenamefont {Horiuchi}\ \emph {et~al.}(2010)\citenamefont
  {Horiuchi}, \citenamefont {Suzuki}, \citenamefont {Capel},\ and\
  \citenamefont {Baye}}]{HOR10}%
  \BibitemOpen
  \bibfield  {author} {\bibinfo {author} {\bibfnamefont {W.}~\bibnamefont
  {Horiuchi}}, \bibinfo {author} {\bibfnamefont {Y.}~\bibnamefont {Suzuki}},
  \bibinfo {author} {\bibfnamefont {P.}~\bibnamefont {Capel}}, \ and\ \bibinfo
  {author} {\bibfnamefont {D.}~\bibnamefont {Baye}},\ }\href {\doibase
  10.1103/PhysRevC.81.024606} {\bibfield  {journal} {\bibinfo  {journal} {Phys.
  Rev. C}\ }\textbf {\bibinfo {volume} {81}},\ \bibinfo {pages} {024606}
  (\bibinfo {year} {2010})}\BibitemShut {NoStop}%
\bibitem [{\citenamefont {Bohr}\ and\ \citenamefont {Mottelson}(1969)}]{BOHR}%
  \BibitemOpen
  \bibfield  {author} {\bibinfo {author} {\bibfnamefont {A.}~\bibnamefont
  {Bohr}}\ and\ \bibinfo {author} {\bibfnamefont {B.~R.}\ \bibnamefont
  {Mottelson}},\ }\href@noop {} {\emph {\bibinfo {title} {Nuclear Structure}}}\
  (\bibinfo  {publisher} {Benjamin, New York},\ \bibinfo {year}
  {1969})\BibitemShut {NoStop}%
\bibitem [{\citenamefont {Kobayashi}\ \emph {et~al.}(2016)\citenamefont
  {Kobayashi}, \citenamefont {Nakamura}, \citenamefont {Kondo}, \citenamefont
  {Tostevin}, \citenamefont {Aoi}, \citenamefont {Baba}, \citenamefont
  {Barthelemy}, \citenamefont {Famiano}, \citenamefont {Fukuda}, \citenamefont
  {Inabe}, \citenamefont {Ishihara}, \citenamefont {Kanungo}, \citenamefont
  {Kim}, \citenamefont {Kubo}, \citenamefont {Lee}, \citenamefont {Lee},
  \citenamefont {Matsushita}, \citenamefont {Motobayashi}, \citenamefont
  {Ohnishi}, \citenamefont {Orr}, \citenamefont {Otsu}, \citenamefont {Sako},
  \citenamefont {Sakurai}, \citenamefont {Satou}, \citenamefont {Sumikama},
  \citenamefont {Takeda}, \citenamefont {Takeuchi}, \citenamefont {Tanaka},
  \citenamefont {Togano},\ and\ \citenamefont {Yoneda}}]{Koba2016}%
  \BibitemOpen
  \bibfield  {author} {\bibinfo {author} {\bibfnamefont {N.}~\bibnamefont
  {Kobayashi}}, \bibinfo {author} {\bibfnamefont {T.}~\bibnamefont {Nakamura}},
  \bibinfo {author} {\bibfnamefont {Y.}~\bibnamefont {Kondo}}, \bibinfo
  {author} {\bibfnamefont {J.~A.}\ \bibnamefont {Tostevin}}, \bibinfo {author}
  {\bibfnamefont {N.}~\bibnamefont {Aoi}}, \bibinfo {author} {\bibfnamefont
  {H.}~\bibnamefont {Baba}}, \bibinfo {author} {\bibfnamefont {R.}~\bibnamefont
  {Barthelemy}}, \bibinfo {author} {\bibfnamefont {M.~A.}\ \bibnamefont
  {Famiano}}, \bibinfo {author} {\bibfnamefont {N.}~\bibnamefont {Fukuda}},
  \bibinfo {author} {\bibfnamefont {N.}~\bibnamefont {Inabe}}, \bibinfo
  {author} {\bibfnamefont {M.}~\bibnamefont {Ishihara}}, \bibinfo {author}
  {\bibfnamefont {R.}~\bibnamefont {Kanungo}}, \bibinfo {author} {\bibfnamefont
  {S.}~\bibnamefont {Kim}}, \bibinfo {author} {\bibfnamefont {T.}~\bibnamefont
  {Kubo}}, \bibinfo {author} {\bibfnamefont {G.~S.}\ \bibnamefont {Lee}},
  \bibinfo {author} {\bibfnamefont {H.~S.}\ \bibnamefont {Lee}}, \bibinfo
  {author} {\bibfnamefont {M.}~\bibnamefont {Matsushita}}, \bibinfo {author}
  {\bibfnamefont {T.}~\bibnamefont {Motobayashi}}, \bibinfo {author}
  {\bibfnamefont {T.}~\bibnamefont {Ohnishi}}, \bibinfo {author} {\bibfnamefont
  {N.~A.}\ \bibnamefont {Orr}}, \bibinfo {author} {\bibfnamefont
  {H.}~\bibnamefont {Otsu}}, \bibinfo {author} {\bibfnamefont {T.}~\bibnamefont
  {Sako}}, \bibinfo {author} {\bibfnamefont {H.}~\bibnamefont {Sakurai}},
  \bibinfo {author} {\bibfnamefont {Y.}~\bibnamefont {Satou}}, \bibinfo
  {author} {\bibfnamefont {T.}~\bibnamefont {Sumikama}}, \bibinfo {author}
  {\bibfnamefont {H.}~\bibnamefont {Takeda}}, \bibinfo {author} {\bibfnamefont
  {S.}~\bibnamefont {Takeuchi}}, \bibinfo {author} {\bibfnamefont
  {R.}~\bibnamefont {Tanaka}}, \bibinfo {author} {\bibfnamefont
  {Y.}~\bibnamefont {Togano}}, \ and\ \bibinfo {author} {\bibfnamefont
  {K.}~\bibnamefont {Yoneda}},\ }\href {\doibase 10.1103/PhysRevC.93.0146R13}
  {\bibfield  {journal} {\bibinfo  {journal} {Phys. Rev. C}\ }\textbf {\bibinfo
  {volume} {93}},\ \bibinfo {pages} {014613} (\bibinfo {year}
  {2016})}\BibitemShut {NoStop}%
\bibitem [{\citenamefont {Baye}(1987{\natexlab{a}})}]{Baye87}%
  \BibitemOpen
  \bibfield  {author} {\bibinfo {author} {\bibfnamefont {D.}~\bibnamefont
  {Baye}},\ }\href {\doibase 10.1103/PhysRevLett.58.2738} {\bibfield  {journal}
  {\bibinfo  {journal} {Phys. Rev. Lett.}\ }\textbf {\bibinfo {volume} {58}},\
  \bibinfo {pages} {2738} (\bibinfo {year} {1987}{\natexlab{a}})}\BibitemShut
  {NoStop}%
\bibitem [{\citenamefont {Baye}(1987{\natexlab{b}})}]{Baye287}%
  \BibitemOpen
  \bibfield  {author} {\bibinfo {author} {\bibfnamefont {D.}~\bibnamefont
  {Baye}},\ }\href {\doibase 10.1088/0305-4470/20/16/027} {\bibfield  {journal}
  {\bibinfo  {journal} {Journal of Physics A: Mathematical and General}\
  }\textbf {\bibinfo {volume} {20}},\ \bibinfo {pages} {5529} (\bibinfo {year}
  {1987}{\natexlab{b}})}\BibitemShut {NoStop}%
\bibitem [{\citenamefont {Gogny}\ \emph {et~al.}(1970)\citenamefont {Gogny},
  \citenamefont {Pires},\ and\ \citenamefont {Tourreil}}]{Gogny70}%
  \BibitemOpen
  \bibfield  {author} {\bibinfo {author} {\bibfnamefont {D.}~\bibnamefont
  {Gogny}}, \bibinfo {author} {\bibfnamefont {P.}~\bibnamefont {Pires}}, \ and\
  \bibinfo {author} {\bibfnamefont {R.~D.}\ \bibnamefont {Tourreil}},\ }\href
  {\doibase https://doi.org/10.1016/0370-2693(70)90552-6} {\bibfield  {journal}
  {\bibinfo  {journal} {Phys. Lett. B}\ }\textbf {\bibinfo {volume} {32}},\
  \bibinfo {pages} {591 } (\bibinfo {year} {1970})}\BibitemShut {NoStop}%
\bibitem [{\citenamefont {Garrido}\ \emph {et~al.}(1997)\citenamefont
  {Garrido}, \citenamefont {Fedorov},\ and\ \citenamefont
  {Jensen}}]{Garrido97}%
  \BibitemOpen
  \bibfield  {author} {\bibinfo {author} {\bibfnamefont {E.}~\bibnamefont
  {Garrido}}, \bibinfo {author} {\bibfnamefont {D.}~\bibnamefont {Fedorov}}, \
  and\ \bibinfo {author} {\bibfnamefont {A.}~\bibnamefont {Jensen}},\ }\href
  {\doibase https://doi.org/10.1016/S0375-9474(97)00044-4} {\bibfield
  {journal} {\bibinfo  {journal} {Nuclear Physics A}\ }\textbf {\bibinfo
  {volume} {617}},\ \bibinfo {pages} {153 } (\bibinfo {year}
  {1997})}\BibitemShut {NoStop}%
\bibitem [{\citenamefont {Garrido}\ \emph {et~al.}(2004)\citenamefont
  {Garrido}, \citenamefont {Fedorov},\ and\ \citenamefont
  {Jensen}}]{Garrido04}%
  \BibitemOpen
  \bibfield  {author} {\bibinfo {author} {\bibfnamefont {E.}~\bibnamefont
  {Garrido}}, \bibinfo {author} {\bibfnamefont {D.~V.}\ \bibnamefont
  {Fedorov}}, \ and\ \bibinfo {author} {\bibfnamefont {A.~S.}\ \bibnamefont
  {Jensen}},\ }\href {\doibase 10.1103/PhysRevC.69.024002} {\bibfield
  {journal} {\bibinfo  {journal} {Phys. Rev. C}\ }\textbf {\bibinfo {volume}
  {69}},\ \bibinfo {pages} {024002} (\bibinfo {year} {2004})}\BibitemShut
  {NoStop}%
\bibitem [{\citenamefont {Ozawa}\ \emph {et~al.}(2001)\citenamefont {Ozawa},
  \citenamefont {Bochkarev}, \citenamefont {Chulkov}, \citenamefont {Cortina},
  \citenamefont {Geissel}, \citenamefont {Hellström}, \citenamefont {Ivanov},
  \citenamefont {Janik}, \citenamefont {Kimura}, \citenamefont {Kobayashi},
  \citenamefont {Korsheninnikov}, \citenamefont {Münzenberg}, \citenamefont
  {Nickel}, \citenamefont {Ogawa}, \citenamefont {Ogloblin}, \citenamefont
  {Pfützner}, \citenamefont {Pribora}, \citenamefont {Simon}, \citenamefont
  {Sitár}, \citenamefont {Strmen}, \citenamefont {Sümmerer}, \citenamefont
  {Suzuki}, \citenamefont {Tanihata}, \citenamefont {Winkler},\ and\
  \citenamefont {Yoshida}}]{Ozawa01}%
  \BibitemOpen
  \bibfield  {author} {\bibinfo {author} {\bibfnamefont {A.}~\bibnamefont
  {Ozawa}}, \bibinfo {author} {\bibfnamefont {O.}~\bibnamefont {Bochkarev}},
  \bibinfo {author} {\bibfnamefont {L.}~\bibnamefont {Chulkov}}, \bibinfo
  {author} {\bibfnamefont {D.}~\bibnamefont {Cortina}}, \bibinfo {author}
  {\bibfnamefont {H.}~\bibnamefont {Geissel}}, \bibinfo {author} {\bibfnamefont
  {M.}~\bibnamefont {Hellström}}, \bibinfo {author} {\bibfnamefont
  {M.}~\bibnamefont {Ivanov}}, \bibinfo {author} {\bibfnamefont
  {R.}~\bibnamefont {Janik}}, \bibinfo {author} {\bibfnamefont
  {K.}~\bibnamefont {Kimura}}, \bibinfo {author} {\bibfnamefont
  {T.}~\bibnamefont {Kobayashi}}, \bibinfo {author} {\bibfnamefont
  {A.}~\bibnamefont {Korsheninnikov}}, \bibinfo {author} {\bibfnamefont
  {G.}~\bibnamefont {Münzenberg}}, \bibinfo {author} {\bibfnamefont
  {F.}~\bibnamefont {Nickel}}, \bibinfo {author} {\bibfnamefont
  {Y.}~\bibnamefont {Ogawa}}, \bibinfo {author} {\bibfnamefont
  {A.}~\bibnamefont {Ogloblin}}, \bibinfo {author} {\bibfnamefont
  {M.}~\bibnamefont {Pfützner}}, \bibinfo {author} {\bibfnamefont
  {V.}~\bibnamefont {Pribora}}, \bibinfo {author} {\bibfnamefont
  {H.}~\bibnamefont {Simon}}, \bibinfo {author} {\bibfnamefont
  {B.}~\bibnamefont {Sitár}}, \bibinfo {author} {\bibfnamefont
  {P.}~\bibnamefont {Strmen}}, \bibinfo {author} {\bibfnamefont
  {K.}~\bibnamefont {Sümmerer}}, \bibinfo {author} {\bibfnamefont
  {T.}~\bibnamefont {Suzuki}}, \bibinfo {author} {\bibfnamefont
  {I.}~\bibnamefont {Tanihata}}, \bibinfo {author} {\bibfnamefont
  {M.}~\bibnamefont {Winkler}}, \ and\ \bibinfo {author} {\bibfnamefont
  {K.}~\bibnamefont {Yoshida}},\ }\href {\doibase
  https://doi.org/10.1016/S0375-9474(01)00563-2} {\bibfield  {journal}
  {\bibinfo  {journal} {Nuclear Physics A}\ }\textbf {\bibinfo {volume}
  {691}},\ \bibinfo {pages} {599 } (\bibinfo {year} {2001})}\BibitemShut
  {NoStop}%
\bibitem [{\citenamefont {Khouaja}\ \emph {et~al.}(2006)\citenamefont
  {Khouaja}, \citenamefont {Villari}, \citenamefont {Benjelloun}, \citenamefont
  {Hirata}, \citenamefont {Auger}, \citenamefont {Savajols}, \citenamefont
  {Mittig}, \citenamefont {Roussel-Chomaz}, \citenamefont {Orr}, \citenamefont
  {Saint-Laurent}, \citenamefont {Pita}, \citenamefont {Gillibert},
  \citenamefont {Chartier}, \citenamefont {Demonchy}, \citenamefont {Giot},
  \citenamefont {Baiborodin}, \citenamefont {Penionzhkevich}, \citenamefont
  {Catford}, \citenamefont {Lépine-Szily},\ and\ \citenamefont
  {Dlouhy}}]{KHO}%
  \BibitemOpen
  \bibfield  {author} {\bibinfo {author} {\bibfnamefont {A.}~\bibnamefont
  {Khouaja}}, \bibinfo {author} {\bibfnamefont {A.}~\bibnamefont {Villari}},
  \bibinfo {author} {\bibfnamefont {M.}~\bibnamefont {Benjelloun}}, \bibinfo
  {author} {\bibfnamefont {D.}~\bibnamefont {Hirata}}, \bibinfo {author}
  {\bibfnamefont {G.}~\bibnamefont {Auger}}, \bibinfo {author} {\bibfnamefont
  {H.}~\bibnamefont {Savajols}}, \bibinfo {author} {\bibfnamefont
  {W.}~\bibnamefont {Mittig}}, \bibinfo {author} {\bibfnamefont
  {P.}~\bibnamefont {Roussel-Chomaz}}, \bibinfo {author} {\bibfnamefont
  {N.}~\bibnamefont {Orr}}, \bibinfo {author} {\bibfnamefont {M.}~\bibnamefont
  {Saint-Laurent}}, \bibinfo {author} {\bibfnamefont {S.}~\bibnamefont {Pita}},
  \bibinfo {author} {\bibfnamefont {A.}~\bibnamefont {Gillibert}}, \bibinfo
  {author} {\bibfnamefont {M.}~\bibnamefont {Chartier}}, \bibinfo {author}
  {\bibfnamefont {C.}~\bibnamefont {Demonchy}}, \bibinfo {author}
  {\bibfnamefont {L.}~\bibnamefont {Giot}}, \bibinfo {author} {\bibfnamefont
  {D.}~\bibnamefont {Baiborodin}}, \bibinfo {author} {\bibfnamefont
  {Y.}~\bibnamefont {Penionzhkevich}}, \bibinfo {author} {\bibfnamefont
  {W.}~\bibnamefont {Catford}}, \bibinfo {author} {\bibfnamefont
  {A.}~\bibnamefont {Lépine-Szily}}, \ and\ \bibinfo {author} {\bibfnamefont
  {Z.}~\bibnamefont {Dlouhy}},\ }\href {\doibase
  https://doi.org/10.1016/j.nuclphysa.2006.07.042} {\bibfield  {journal}
  {\bibinfo  {journal} {Nucl. Phys. A}\ }\textbf {\bibinfo {volume} {780}},\
  \bibinfo {pages} {1 } (\bibinfo {year} {2006})}\BibitemShut {NoStop}%
\bibitem [{\citenamefont {Raynal}\ and\ \citenamefont {Revai}(1970)}]{RR70}%
  \BibitemOpen
  \bibfield  {author} {\bibinfo {author} {\bibfnamefont {J.}~\bibnamefont
  {Raynal}}\ and\ \bibinfo {author} {\bibfnamefont {J.}~\bibnamefont {Revai}},\
  }\href {\doibase 10.1007/BF02756127} {\bibfield  {journal} {\bibinfo
  {journal} {Nuovo Cim.}\ }\textbf {\bibinfo {volume} {68A}},\ \bibinfo {pages}
  {612} (\bibinfo {year} {1970})}\BibitemShut {NoStop}%
\bibitem [{\citenamefont {Catara}\ \emph {et~al.}(1984)\citenamefont {Catara},
  \citenamefont {Insolia}, \citenamefont {Maglione},\ and\ \citenamefont
  {Vitturi}}]{Catara84}%
  \BibitemOpen
  \bibfield  {author} {\bibinfo {author} {\bibfnamefont {F.}~\bibnamefont
  {Catara}}, \bibinfo {author} {\bibfnamefont {A.}~\bibnamefont {Insolia}},
  \bibinfo {author} {\bibfnamefont {E.}~\bibnamefont {Maglione}}, \ and\
  \bibinfo {author} {\bibfnamefont {A.}~\bibnamefont {Vitturi}},\ }\href
  {\doibase 10.1103/PhysRevC.29.1091} {\bibfield  {journal} {\bibinfo
  {journal} {Phys. Rev. C}\ }\textbf {\bibinfo {volume} {29}},\ \bibinfo
  {pages} {1091} (\bibinfo {year} {1984})}\BibitemShut {NoStop}%
\bibitem [{\citenamefont {Hagino}\ and\ \citenamefont {Sagawa}(2005)}]{sag15}%
  \BibitemOpen
  \bibfield  {author} {\bibinfo {author} {\bibfnamefont {K.}~\bibnamefont
  {Hagino}}\ and\ \bibinfo {author} {\bibfnamefont {H.}~\bibnamefont
  {Sagawa}},\ }\href {\doibase 10.1103/PhysRevC.72.044321} {\bibfield
  {journal} {\bibinfo  {journal} {Phys. Rev. C}\ }\textbf {\bibinfo {volume}
  {72}},\ \bibinfo {pages} {044321} (\bibinfo {year} {2005})}\BibitemShut
  {NoStop}%
\bibitem [{\citenamefont {Corsi}\ \emph {et~al.}(2019)\citenamefont {Corsi},
  \citenamefont {Kubota}, \citenamefont {Casal}, \citenamefont {Gómez-Ramos},
  \citenamefont {Moro}, \citenamefont {Authelet}, \citenamefont {Baba},
  \citenamefont {Caesar}, \citenamefont {Calvet}, \citenamefont {Delbart},
  \citenamefont {Dozono}, \citenamefont {Feng}, \citenamefont {Flavigny},
  \citenamefont {Gheller}, \citenamefont {Gibelin}, \citenamefont {Giganon},
  \citenamefont {Gillibert}, \citenamefont {Hasegawa}, \citenamefont {Isobe},
  \citenamefont {Kanaya}, \citenamefont {Kawakami}, \citenamefont {Kim},
  \citenamefont {Kiyokawa}, \citenamefont {Kobayashi}, \citenamefont
  {Kobayashi}, \citenamefont {Kobayashi}, \citenamefont {Kondo}, \citenamefont
  {Korkulu}, \citenamefont {Koyama}, \citenamefont {Lapoux}, \citenamefont
  {Maeda}, \citenamefont {Marqués}, \citenamefont {Motobayashi}, \citenamefont
  {Miyazaki}, \citenamefont {Nakamura}, \citenamefont {Nakatsuka},
  \citenamefont {Nishio}, \citenamefont {Obertelli}, \citenamefont {Ohkura},
  \citenamefont {Orr}, \citenamefont {Ota}, \citenamefont {Otsu}, \citenamefont
  {Ozaki}, \citenamefont {Panin}, \citenamefont {Paschalis}, \citenamefont
  {Pollacco}, \citenamefont {Reichert}, \citenamefont {Rousse}, \citenamefont
  {Saito}, \citenamefont {Sakaguchi}, \citenamefont {Sako}, \citenamefont
  {Santamaria}, \citenamefont {Sasano}, \citenamefont {Sato}, \citenamefont
  {Shikata}, \citenamefont {Shimizu}, \citenamefont {Shindo}, \citenamefont
  {Stuhl}, \citenamefont {Sumikama}, \citenamefont {Sun}, \citenamefont
  {Tabata}, \citenamefont {Togano}, \citenamefont {Tsubota}, \citenamefont
  {Uesaka}, \citenamefont {Yang}, \citenamefont {Yasuda}, \citenamefont
  {Yoneda},\ and\ \citenamefont {Zenihiro}}]{corsi19}%
  \BibitemOpen
  \bibfield  {author} {\bibinfo {author} {\bibfnamefont {A.}~\bibnamefont
  {Corsi}}, \bibinfo {author} {\bibfnamefont {Y.}~\bibnamefont {Kubota}},
  \bibinfo {author} {\bibfnamefont {J.}~\bibnamefont {Casal}}, \bibinfo
  {author} {\bibfnamefont {M.}~\bibnamefont {Gómez-Ramos}}, \bibinfo {author}
  {\bibfnamefont {A.}~\bibnamefont {Moro}}, \bibinfo {author} {\bibfnamefont
  {G.}~\bibnamefont {Authelet}}, \bibinfo {author} {\bibfnamefont
  {H.}~\bibnamefont {Baba}}, \bibinfo {author} {\bibfnamefont {C.}~\bibnamefont
  {Caesar}}, \bibinfo {author} {\bibfnamefont {D.}~\bibnamefont {Calvet}},
  \bibinfo {author} {\bibfnamefont {A.}~\bibnamefont {Delbart}}, \bibinfo
  {author} {\bibfnamefont {M.}~\bibnamefont {Dozono}}, \bibinfo {author}
  {\bibfnamefont {J.}~\bibnamefont {Feng}}, \bibinfo {author} {\bibfnamefont
  {F.}~\bibnamefont {Flavigny}}, \bibinfo {author} {\bibfnamefont {J.-M.}\
  \bibnamefont {Gheller}}, \bibinfo {author} {\bibfnamefont {J.}~\bibnamefont
  {Gibelin}}, \bibinfo {author} {\bibfnamefont {A.}~\bibnamefont {Giganon}},
  \bibinfo {author} {\bibfnamefont {A.}~\bibnamefont {Gillibert}}, \bibinfo
  {author} {\bibfnamefont {K.}~\bibnamefont {Hasegawa}}, \bibinfo {author}
  {\bibfnamefont {T.}~\bibnamefont {Isobe}}, \bibinfo {author} {\bibfnamefont
  {Y.}~\bibnamefont {Kanaya}}, \bibinfo {author} {\bibfnamefont
  {S.}~\bibnamefont {Kawakami}}, \bibinfo {author} {\bibfnamefont
  {D.}~\bibnamefont {Kim}}, \bibinfo {author} {\bibfnamefont {Y.}~\bibnamefont
  {Kiyokawa}}, \bibinfo {author} {\bibfnamefont {M.}~\bibnamefont {Kobayashi}},
  \bibinfo {author} {\bibfnamefont {N.}~\bibnamefont {Kobayashi}}, \bibinfo
  {author} {\bibfnamefont {T.}~\bibnamefont {Kobayashi}}, \bibinfo {author}
  {\bibfnamefont {Y.}~\bibnamefont {Kondo}}, \bibinfo {author} {\bibfnamefont
  {Z.}~\bibnamefont {Korkulu}}, \bibinfo {author} {\bibfnamefont
  {S.}~\bibnamefont {Koyama}}, \bibinfo {author} {\bibfnamefont
  {V.}~\bibnamefont {Lapoux}}, \bibinfo {author} {\bibfnamefont
  {Y.}~\bibnamefont {Maeda}}, \bibinfo {author} {\bibfnamefont
  {F.}~\bibnamefont {Marqués}}, \bibinfo {author} {\bibfnamefont
  {T.}~\bibnamefont {Motobayashi}}, \bibinfo {author} {\bibfnamefont
  {T.}~\bibnamefont {Miyazaki}}, \bibinfo {author} {\bibfnamefont
  {T.}~\bibnamefont {Nakamura}}, \bibinfo {author} {\bibfnamefont
  {N.}~\bibnamefont {Nakatsuka}}, \bibinfo {author} {\bibfnamefont
  {Y.}~\bibnamefont {Nishio}}, \bibinfo {author} {\bibfnamefont
  {A.}~\bibnamefont {Obertelli}}, \bibinfo {author} {\bibfnamefont
  {A.}~\bibnamefont {Ohkura}}, \bibinfo {author} {\bibfnamefont
  {N.}~\bibnamefont {Orr}}, \bibinfo {author} {\bibfnamefont {S.}~\bibnamefont
  {Ota}}, \bibinfo {author} {\bibfnamefont {H.}~\bibnamefont {Otsu}}, \bibinfo
  {author} {\bibfnamefont {T.}~\bibnamefont {Ozaki}}, \bibinfo {author}
  {\bibfnamefont {V.}~\bibnamefont {Panin}}, \bibinfo {author} {\bibfnamefont
  {S.}~\bibnamefont {Paschalis}}, \bibinfo {author} {\bibfnamefont
  {E.}~\bibnamefont {Pollacco}}, \bibinfo {author} {\bibfnamefont
  {S.}~\bibnamefont {Reichert}}, \bibinfo {author} {\bibfnamefont {J.-Y.}\
  \bibnamefont {Rousse}}, \bibinfo {author} {\bibfnamefont {A.}~\bibnamefont
  {Saito}}, \bibinfo {author} {\bibfnamefont {S.}~\bibnamefont {Sakaguchi}},
  \bibinfo {author} {\bibfnamefont {M.}~\bibnamefont {Sako}}, \bibinfo {author}
  {\bibfnamefont {C.}~\bibnamefont {Santamaria}}, \bibinfo {author}
  {\bibfnamefont {M.}~\bibnamefont {Sasano}}, \bibinfo {author} {\bibfnamefont
  {H.}~\bibnamefont {Sato}}, \bibinfo {author} {\bibfnamefont {M.}~\bibnamefont
  {Shikata}}, \bibinfo {author} {\bibfnamefont {Y.}~\bibnamefont {Shimizu}},
  \bibinfo {author} {\bibfnamefont {Y.}~\bibnamefont {Shindo}}, \bibinfo
  {author} {\bibfnamefont {L.}~\bibnamefont {Stuhl}}, \bibinfo {author}
  {\bibfnamefont {T.}~\bibnamefont {Sumikama}}, \bibinfo {author}
  {\bibfnamefont {Y.}~\bibnamefont {Sun}}, \bibinfo {author} {\bibfnamefont
  {M.}~\bibnamefont {Tabata}}, \bibinfo {author} {\bibfnamefont
  {Y.}~\bibnamefont {Togano}}, \bibinfo {author} {\bibfnamefont
  {J.}~\bibnamefont {Tsubota}}, \bibinfo {author} {\bibfnamefont
  {T.}~\bibnamefont {Uesaka}}, \bibinfo {author} {\bibfnamefont
  {Z.}~\bibnamefont {Yang}}, \bibinfo {author} {\bibfnamefont {J.}~\bibnamefont
  {Yasuda}}, \bibinfo {author} {\bibfnamefont {K.}~\bibnamefont {Yoneda}}, \
  and\ \bibinfo {author} {\bibfnamefont {J.}~\bibnamefont {Zenihiro}},\ }\href
  {\doibase https://doi.org/10.1016/j.physletb.2019.134843} {\bibfield
  {journal} {\bibinfo  {journal} {Phys. Lett. B}\ }\textbf {\bibinfo {volume}
  {797}},\ \bibinfo {pages} {134843} (\bibinfo {year} {2019})}\BibitemShut
  {NoStop}%
\end{thebibliography}%

\end{document}